\documentclass[a4paper]{article}

\usepackage{jheppub} 
                     
\usepackage{graphicx,color}
\usepackage[export]{adjustbox} 
 \usepackage{bm}
   \usepackage{amsmath}
    \usepackage{amssymb}    
     \usepackage{pifont}
 \usepackage{stmaryrd}

\usepackage{standalone}
\usepackage{slashed}
\usepackage{multirow}
\usepackage{makecell}
\usepackage{boldline}
\usepackage{tabu}
\usepackage{hhline}
\usepackage{colortbl}
\usepackage{gensymb}

\newcommand{\nn}{\nonumber}
\newcommand{\df}{\mathrm{d}}

\renewcommand{\(}{\left(}
\renewcommand{\)}{\right)}

\renewcommand{\vec}[1]{\bm{#1}}

\title{
Assessing the sensitivity of Energy–Energy Correlations in $e^+e^-$ annihilation to TMD dynamics
}

\author{Alejandro Bris Cuerpo,}
\author{Ignazio Scimemi,}
\author{Alexey Vladimirov}

\affiliation{
Departamento de F\'isica Te\'orica \& IPARCOS, Facultad de Ciencias Físicas, Universidad Complutense de Madrid, Plaza de Ciencias 1, E-28040 Madrid, Spain}

\emailAdd{albris@ucm.es}
\emailAdd{ignazios@ucm.es}
\emailAdd{alexeyvl@ucm.es}

\preprint{IPARCOS-UCM-25-028}

\abstract{
We critically examine the back-to-back limit of the energy–energy correlation (EEC) in $e^+e^-$ annihilation as a potential source of information on the Collins–Soper kernel and the strong coupling constant. The analysis is performed within the framework of transverse momentum dependent (TMD) factorization at next-to-next-to-next-to-next-to-leading logarithmic (N$^4$LL) accuracy, using a global fit to all available experimental data. Contrary to previous claims, we demonstrate that the current data do not provide meaningful constraints on either the Collins–Soper kernel or $\alpha_s$.
}

\begin{document} 
\allowdisplaybreaks
\maketitle 
\section{Introduction}
For a long time, the energy-energy correlation (EEC) in two-particle leptonic production, $e^+e^- \rightarrow ij+X$, has been recognized as a relevant observable for studying both perturbative and nonperturbative aspects of quantum chromodynamics (QCD). The EEC is an infrared-safe event shape, defined in  ref.~\cite{Basham:1978bw,Basham:1978zq} as
\begin{align}
\label{eq:EECdef}
 \text{EEC}(\chi) =  \frac{\df \sigma}{\df \chi}= \sum_{i,j} \int \df \sigma_{e^+e^-\to i j + X} \, \frac{E_i E_j}{Q^2} \, \delta(\cos\theta_{ij} - \cos \chi)
\,,\end{align}
where the sum runs over all pairs of produced particles $\{i, j\}$ with energies $E_i$ and $E_j$ and the polar angle $\theta_{ij}$ between their momenta. The variable $Q^2$ is the invariant mass of the $e^+ e^-$ collision. The theoretical description of EEC is different depending on angle. So, in the mid-angle regime ($\chi \sim 90^o$), EEC is entirely perturbative and depends only on the strong coupling constant $\alpha_s$. Therefore, the mid-angle data for EEC is refereed as one of the best sources for the determination of the strong coupling constants $\alpha_s$, with the first extractions made already at LEP and SLAC \cite{OPAL:1990reb, SLD:1994yoe, SLD:1994idb, DELPHI:1990sof, DELPHI:1993nlw, OPAL:1993pnw}. For $\chi$ approaching collinear ($\chi\sim 0^o$) or back-to-back ($\chi\sim 180^o$) limits the nonperturbative QCD effects should be taken into account.

In this work, we study the back-to-back limit of the EEC. In this regime, the EEC is described by the TMD factorization approach~\cite{Moult:2018jzp,Ebert:2020sfi}, which accounts for soft corrections to parton momenta that cannot be described by perturbative computations. As a consequence, the theoretical description includes additional nonperturbative parameters beyond $\alpha_s$. Investigating these nonperturbative effects can provide insight into the structure of the QCD vacuum~\cite{Vladimirov:2020umg} and the confinement/deconfinement transition~\cite{Lee:2024esz,Lee:2025okn}. In the present study, we focus on the Collins–Soper (CS) kernel, a universal nonperturbative function that governs TMD evolution.

The CS kernel is one of the most fundamental QCD quantities which is used to describe a multitude of processes and data and for this reason, it attracts a lot of attention. It has been determined from the data for Drell-Yan and Semi-Inclusive Deep-Inelastic Scattering (for recent determinations see \cite{Bacchetta:2019sam, Bertone:2019nxa, Scimemi:2019cmh, Bacchetta:2022awv, Moos:2023yfa, Bacchetta:2024qre, Moos:2025sal}, and also via the QCD lattice simulations (for recent determinations see \cite{Schlemmer:2021aij, Shu:2023cot, Avkhadiev:2023poz, Avkhadiev:2024mgd, Bollweg:2024zet, Bollweg:2025iol}). The perturbative asymptotic of CS kernel is known up to third order (N$^3$LO) (four-loops) \cite{Li:2016ctv, Vladimirov:2017ksc, Duhr:2022yyp, Moult:2022xzt}, including also estimations for following power correction \cite{Korchemsky:1995zm, Scimemi:2016ffw, Vladimirov:2020umg}. Even so, the available curves for CS kernel are rather unprecise for $b>$1.-1.5GeV$^{-1}$. Given the high precision of EEC data, one can hope that these data provide a decisive constraint to the CS kernel, alike it does for $\alpha_s$.

An analysis of EEC data within the TMD factorization framework was recently conducted in Ref.~\cite{Kang:2024dja}. It was concluded that EEC data are indeed capable of constraining the CS kernel alongside $\alpha_s$ and the nonperturbative parameters of the jet functions. These results merit further study to assess whether the precision of the extracted functions is compatible with current knowledge.

The goal of this work is to examine global EEC data and determine their constraining power with respect to the CS kernel and other QCD parameters. The analysis is based on the current state-of-the-art implementation of the TMD factorization theorem, incorporating the maximum available perturbative information (denoted briefly as N$^4$LL). This formulation was recently used in more traditional data fits~\cite{Moos:2023yfa,Moos:2025sal}, which serve as a natural reference point for values and precision. The analysis is also based on the same code, \texttt{artemide}~\cite{artemide, Scimemi:2017etj}, which makes it an ideal framework for this study.

The presentation is structured as following: we start from the practical review of the theory in sec.~\ref{sec:theory}. Then in section \ref{sec:data} we describe the data selection criterion and overview available data for EEC in the back-to-back regime. Finally, in sec.~\ref{sec:stat} and \ref{sec:fit} we present the fitting procedure and its results, emphasizing the issues in the uncertainty propagation that we faced in this work. The conclusions are reported in sec.~\ref{sec:conclusions}.

\section{Notation and formalism}
\label{sec:theory}

The EEC in $e^+e^-$ is commonly described in  terms of the kinematic variable $z$ defined as:
\begin{align}
z=\frac{1-\cos(\chi)}{2}\,,
\end{align}
where $\chi$ is the angle between the particles. The back-to-back limit ($\chi \to 180^{\circ}$), corresponds to $z\to 1$. In this limit, the energy-energy correlation can be factorized at leading power of $(1-z)Q$ , using the TMD factorization formalism \cite{Ebert:2020sfi}. The resulting expression for the cross-section is
\begin{align}
\label{eq:Xsec}
\frac{{\rm d}\sigma }{{\rm d}z}=\frac{\hat{\sigma}_0}{8}\sum_{f,\bar f}H_{f\bar{f}}(Q,\mu)\!\int_{0}^{\infty}{\rm d}(b_TQ)^2\,J_0(b_TQ\sqrt{1-z})J_f(b_T,\mu,\zeta)J_{\bar{f}}(b_T,\mu,\bar\zeta)\,,
\end{align}
where $f,\;\bar f=q,\bar q$ and $\zeta\bar \zeta=Q^4$. Notice that in this expression the variable $b_T$ is the conjugate variable of $Q\sqrt{1-z}$. In eq.~(\ref{eq:Xsec}), $\hat{\sigma}_0$ is the Born level cross section, $H_{f\bar{f}}(Q,\mu)$ is the hard factor, $J_0$ is the Bessel function of the first kind, and  $J_{f(\bar f)}$ is the jet function. The hard factor $H$ is the same as in the Drell-Yan reaction, and is know up to N$^4$LO \cite{Lee:2022nhh}.

The energy-jet function is defined as a production of an any-type hadron by any collinear momentum fraction. Thus it is related to the unpolarized TMD fragmentation function (TMDFF) as
\begin{align}\label{eq.Jet-TMD}
J_{f}\(b_T;\mu,\zeta\) &\equiv  \sum_{h} \int_0^1 \df z   \,z^3  D_{f\rightarrow h}\(z, b_T;\mu,\zeta\)\,.
\end{align}
The variables $\mu$ and $\zeta$ are ordinary ultra-violet and rapidity factorization scales of a TMD distribution. Since the TMD evolution is independent on $z$, the evolution equations for the jet function coincides with those for the TMDFF. They are
\begin{eqnarray}\label{def:evol}
\frac{d}{d\ln\mu^2}\ln J(b_T;\mu,\zeta)=\frac{\Gamma_{\text{cusp}}(\mu)}{2}\ln\(\frac{\mu^2}{\zeta}\)-\frac{\gamma_V(\mu)}{2},
\qquad
\frac{d}{d\ln\zeta}\ln J(b_T;\mu,\zeta)=-\mathcal{D}(b_T,\mu),
\end{eqnarray}
where $\Gamma_{\text{cusp}}$ is the cusp anomalous dimension, $\gamma_V$ is the light-like-quark anomalous dimension, and $\mathcal{D}$ is the Collins-Soper kernel.  In the present work, we use the best available precision for evolution kernels, as so, the $\Gamma_{\text{cusp}}$ at N$^4$LO \cite{Moch:2018wjh, Herzog:2018kwj}, and $\gamma_V$ at N$^3$LO \cite{Lee:2022nhh}.

The evolution part with respect to $\mu$ and $\zeta$ can be carried out of the jet function with the help of the $\zeta$-prescription~\cite{Scimemi:2018xaf}. In this case, the cross-section takes a simpler form
\begin{align}
\label{eq:Xsec1}
\frac{{\rm d}\sigma }{{\rm d}z}=\frac{\hat{\sigma}_0}{8}\sum_{f,\bar f}H_{f\bar{f}}(Q,\mu)\!\int_{0}^{\infty}{\rm d}(b_TQ)^2\,J_0(b_TQ\sqrt{1-z})R(\mu,Q)J_f(b_T)J_{\bar{f}}(b_T)\,.
\end{align}
with 
\begin{align}
R(\mu,Q)=\left(\frac{Q^2}{\zeta_Q(b_T^2)}\right)^{-2{\cal D}(b_T^2,\mu)}   
\end{align}
and $\zeta_Q$ is the equi-evolution line, which passes though the saddle point of evolution field \cite{Scimemi:2018xaf, Vladimirov:2019bfa}. Note, that in eq.~(\ref{eq:Xsec1}) the jet functions are independent on scales $\mu$ and $\zeta$ since they are exactly scale-independent. For this reason, they are commonly called as optimal jet-functions.

\noindent Experimental measurements are often provided as normalized to the total cross-section $\sigma_{t}$
\begin{align}\label{eq:EECdists}
\frac{1}{\sigma_{t}}\frac{{\rm d}\sigma}{{\rm d} z}
~~
\text{ or }
~~
\frac{1}{\sigma_{t}}\frac{{\rm d}\sigma}{{\rm d} \chi}\overset{\left[\text{rad}^{-1}\right]}{=}\frac{\sin\chi}{2\,\sigma_{t}}\frac{{\rm d}\sigma}{{\rm d} z} 
~~
\text{ or } 
~~
\frac{1}{\sigma_{t}}\frac{{\rm d}\sigma}{{\rm d} \cos\chi}=\frac{1}{2\,\sigma_{t}}\frac{{\rm d}\sigma}{{\rm d} z}.
\end{align}
In these cases, the partonic cross-section in eq.~(\ref{eq:Xsec}) is replaced by \cite{Tulipant:2017ybb}
\begin{align}\label{normalization-factor}
\frac{\hat{\sigma}_0}{\sigma_t}&=1-a_s A_t+a_s^2 (A_t^2-B_t)+..
\\\nn 
A_t&=3C_F,\qquad B_t=4C_F\left[\left(\frac{123}{8}-11\zeta_3\right)C_A-\frac{3}{8}C_F-\left(\frac{11}{2}-4\zeta_3\right)n_fT_R\right].
\end{align}
Note, that this normalization does not provide a sufficient precision and disagrees with the data at the level of 3-5\% at $Q\sim M_Z$. This can be described by large power corrections and renormalons \cite{Schindler:2023cww}. Furthermore, as we demonstrate below, some data sets disagrees with each other in the normalization, therefore, for the fit of the data we finetune the normalization factor to the data, as described in sec.~\ref{sec:stat}.

It is also important to mention that due to the significant precision of the data, one should take into account the finite-bin effects. It means that the differential cross-section for each bin must be integrated within the bin-limits and divided by the  bin-size.

\subsection{Collins-Soper kernel}

The Collins-Soper kernel is a non-perturbative function. Here we use the same model as in the phenomenological analyses ART23~\cite{Moos:2023yfa} and ART25~\cite{Moos:2025sal}. This model reads
\begin{eqnarray}\label{def:CS-kernel}
{\cal D}(b,\mu)={\cal D}_{\text{pert}}(b^*,\mu^*)+\int_{\mu^*}^\mu \frac{d\mu'}{\mu'}\Gamma_{\text{cusp}}(\mu')+{\cal D}_{\text{NP}}(b),
\end{eqnarray}
where
\begin{eqnarray}\label{CS:scale}
b^*(b)=\frac{b}{\sqrt{1+\frac{\vec b^2}{B^2_{\text{NP}}}}},\qquad \mu^*(b)=\frac{2e^{-\gamma_E}}{b^*(b)}\,
.
\end{eqnarray}
Here, ${\cal D}_{\text{pert}}$ is the perturbative expression of the CS kernel at N$^3$LO (i.e. including terms $\sim a_s^4$ \cite{Duhr:2022yyp, Moult:2022xzt}), and ${\cal D}_{\text{NP}}$ accumulates the effects of power and non-perturbative corrections. The model for ${\cal D}_{\text{NP}}$ is
\begin{align}
\label{CS:NP-part}
{\cal{D}}_{\text{NP}}(b)=bb^*\left[c_0+c_1\ln \left(\frac{b^*}{B_{\text{NP}}}\right)\right],
\end{align}
with $c_0$ and $c_1$ free parameters.

The parameter $B_{\text{NP}}$ was determined in ART23 from the fit of the data as $B_{\text{NP}}=1.56_{-0.09}^{+0.13}$GeV. For that reason, it is fixed to the value 
$$B_{\text{NP}}=1.5\text{ GeV},$$
in ART25 and in the present work.

The parameters $c_0$ and $c_1$ were determined as
\begin{eqnarray}\label{CS:ART23}
\text{ART23~~\cite{Moos:2023yfa}:} &\qquad & 
c_0=0.0369^{+0.0069}_{-0.0061},
\qquad 
c_1=0.0582^{+0.0064}_{-0.0088},
\\\label{CS:ART25}
\text{ART25~~\cite{Moos:2025sal}:} &\qquad & 
c_0=0.0859^{+0.0023}_{-0.0017},
\qquad 
c_1=0.0303^{+0.0038}_{-0.0041}.
\end{eqnarray}
These parameters have significant correlation ($\sim 0.8$). Clearly, the values for $c_{0,1}$ determined in ART23 and ART25 are inconsistent, despite they were determined in identical frameworks. The only difference is the fit ART23 was based on the data for the Drell-Yan reaction only, while ART25 used the same together with SIDIS data. It implies that the uncertainty on these parameters are seriously underestimated, or that they are biased by the model for CS kernel. Note, that other studies (phenomenological or lattice) do not resolve this issue, and they also provide a rather large spread of predictions. For a review and collection of recent results for CS kernel see ref.~\cite{Moos:2025sal}.

\subsection{The jet functions}

The  non-perturbative (optimal) jet function ansatz can be constructed in the same manner as ansatzes for other TMD distributions. This ansatz has the following form
\begin{align}\label{def:J-f}
J_f(b_T)=J^{\text{pert.}}(b_T)J_f^{\text{NP}}(b_T),
\end{align}
where $J^{\text{pert.}}$ is the perturbative approximation of the jet function, which is valid in the regime of small-$b_T$. The $J_f^{\text{NP}}$ is the correction to the perturbative part that must turn to 1 at $b_T\to 0$.

The perturbative part of the jet function can be derived from the definition eq.~(\ref{eq.Jet-TMD}). The small-$b$ asymptotic of TMDFF is given by
\begin{align}
D_{f\rightarrow h}\left(z, b_T\right)=\frac{1}{z^2}\int_z^1\frac{\df y}{y}\,\sum_{f'}y^2\mathbb{C}_{f\to f'}\left(y,b,\mu_{\text{OPE}}\right)d_{1,f'\to h}\left(\frac{z}{y},\mu_{\text{OPE}}\right),
\end{align}
where $\mathbb{C}$ is the perturbative coefficient function known at N$^3$LO \cite{Echevarria:2015usa, Echevarria:2016scs, Ebert:2020qef, Luo:2020epw}. The Mellin moment can be easily computed using the sum rule for unpolarized FF, 
\begin{align} \label{eq:sum_rule1}
 \sum_h \int_0^1 \df z \, z \, d_{f\rightarrow h}(z,\mu) = 1
\,.
\end{align}
It leads to the following expression
\begin{eqnarray}\label{def:D-integrated}
\mathbb{J}(b_T;\mu_{\text{OPE}})&=&
\sum_{f'} \int_0^1\df y\,y^3\mathbb{C}_{f\to f'}\left(y,b,\mu_{\text{OPE}}\right)
=
\sum_{n=0}^\infty a^n_s(\mu_{\text{OPE}})\sum_{k=0}^n \mathbf{L}_b^k \mathbb{J}_{nk},
\end{eqnarray}
where $a_s=g^2/(4\pi)^2=\alpha_s/(4\pi)$ and 
\begin{eqnarray}
\mathbf{L}_b=\ln\(\frac{\mu_{\text{OPE}}^2 \vec b^2}{4e^{-2\gamma_E}}\),
\end{eqnarray}
and $\mathbb{J}_{nk}$ are perturbative coefficients. Up to three loop order they read
\begin{eqnarray}\label{def:coef-perturbative}
\mathbb{J}_{00}&=&1,
\\\nn
\mathbb{J}_{11}&=&0,\qquad \mathbb{J}_{10}=C_F\(4-9\zeta_2\),
\\\nn
\mathbb{J}_{22}&=&0,\qquad \mathbb{J}_{21}=\beta_0 \mathbb{J}_{10},
\\\nn
\mathbb{J}_{20}&=&C_F\Bigg\{
C_F\(\frac{139}{24}-32\zeta_2-74\zeta_3+\frac{645}{4}\zeta_4\)
+C_A\(\frac{9109}{648}-\frac{141}{2}\zeta_2+\frac{334}{{9}}\zeta_3\)+
\\\nn && \qquad N_f\(-\frac{989}{324}+11\zeta_2+\frac{26}{9}\zeta_3\)
\Bigg\},
\\\nn
\mathbb{J}_{33}&=&0,\qquad \mathbb{J}_{32}=\beta_0^2\mathbb{J}_{10},\qquad \mathbb{J}_{31}=2\beta_0 \mathbb{J}_{20}+\beta_1 \mathbb{J}_{10},
\end{eqnarray}
\begin{eqnarray}
&&\mathbb{J}_{30}=
C_F \Bigg\{C_F^2\(\frac{163}{4}+\frac{37}{24}\zeta_2+\frac{32}{3}\zeta_3+318\zeta_4+674\zeta_2\zeta_3-\frac{496}{3}\zeta_3^2+\frac{2008}{3}\zeta_5-\frac{310897}{144}\zeta_6\)
\\\nn &&\qquad
+C_FC_A\(\frac{47419}{648}-\frac{{100439}}{216}\zeta_2-\frac{94889}{54}\zeta_3-\frac{2842}{9}\zeta_2\zeta_3+\frac{688}{3}\zeta_3^2+\frac{230195}{108}\zeta_4+\frac{{9620}}{9}\zeta_5+\frac{1625}{9}\zeta_6\)
\\\nn &&\qquad
+C^2_A\(\frac{3796055}{52488}-\frac{681797}{729}\zeta_2+\frac{169150}{{243}}\zeta_3+\frac{2431}{9}\zeta_2\zeta_3-\frac{565}{9}\zeta_3^2+\frac{31351}{108}\zeta_4-\frac{9022}{9}\zeta_5-\frac{799}{27}\zeta_6\)
\\\nn &&\qquad
+C_FN_f\(-\frac{14623}{243}+\frac{16369}{108}\zeta_2+\frac{21088}{81}\zeta_3-\frac{674}{9}\zeta_2\zeta_3-\frac{17665}{54}\zeta_4-\frac{1072}{9}\zeta_5\)
\\\nn &&\qquad
+C_AN_f\(-\frac{379735}{13122}+\frac{190013}{729}\zeta_2+\frac{2023}{81}\zeta_3+\frac{14}{9}\zeta_2\zeta_3-\frac{560}{27}\zeta_4+\frac{244}{3}\zeta_5\)
\\\nn &&\qquad
+N^2_f\(\frac{22811}{13122}-\frac{388}{27}\zeta_2-\frac{4936}{243}\zeta_3-\frac{134}{27}\zeta_4\)\Bigg\},
\end{eqnarray}
where $\beta_n$ are the coefficients of QCD $\beta$-function, $C_A$ and $C_F$ are usual Casimir eigenvalues, and $N_f$ is the number of active quark flavors. Note, that this function is independent on the flavor of the quark that initiate the jet.

The dependence on the scale $\mu_{\text{OPE}}$ for TMDFFs cancels between the FFs and the coefficient function. In the integrated expression eq.~(\ref{def:D-integrated}) the dependence on $\mu_{\text{OPE}}$ cancels only with the scale-dependence of $\alpha_s$. Therefore, all logarithms of eq.~(\ref{def:D-integrated}) can be absorbed into the argument of $\alpha_s$, which becomes $\mu_{\text{OPE}}\to 2e^{-\gamma_E}/b$. 

At large $b$ the expression (\ref{def:D-integrated}) is not valid since it receives power corrections. Simultaneously, the perturbative expansion becomes unstable, due to $\mu_{\text{OPE}}$ approaching the Landau pole. To avoid this problem, we freeze the perturbative part to a certain value by setting
\begin{eqnarray}\label{mu:OPE}
\mu_{\text{OPE}}=\frac{2 e^{-\gamma_E}}{b^*_J},\qquad \text{with}\qquad b^*_J(b)=\frac{b}{\sqrt{1+\frac{\vec b^2}{B^{*2}_{\text{J}}}}}.
\end{eqnarray}
We fix the parameter $B^{*}_J=0.2$ GeV$^{-1}$, which roughly correspond to the offset of 5 GeV used in $\mu_{\text{OPE}}$ in ART23 and ART25 fits. These values are selected such that our energy scales do not cross the heavy quark mass thresholds, i.e. $\mu_{\text{OPE}}>m_b$. In this way, we avoid the problem of matching between perturbative expressions at thresholds and use $N_f=5$ everywhere.  For the discussion of quark mass contribution to EEC see refs.~\cite{Aglietti:2024xwv, Aglietti:2024zhg}.

Thus our final expression for the perturbative component of the jet function is
\begin{eqnarray}
J^{\text{pert.}}(b_T)=\sum_{n=0}^\infty \mathbb{J}_{n0}a^n_s\(\frac{2 e^{-\gamma_E}}{b^*_J}\).
\end{eqnarray}
At small-$b$ this expression exactly reproduces the perturbative asymptotic, and at large $b$ it is a constant.

The non-perturbative ansatz is a subject of modeling. Conceptually, it could be derived from the non-perturbative parts of TMDFFs. However, practically, such approach fails, because the sum rule eq.~(\ref{eq:sum_rule1}) requires the summation of all available hadrons whose TMD parts are unknown. Moreover, superficial studies show that the distributions of known hadrons (pions, kaons, and proton) do not fulfill the sum rule. Therefore, one should consider them as independent functions.

In what follows, we study two models. The model 1 has a single parameter $a_1$
\begin{eqnarray}\label{eq:m1}
\text{Model 1~:}\qquad J^{\text{NP}}(b_T)=e^{-a_1 b_T}\,,
\end{eqnarray}
and the model 2 has three parameters $a_{1,2,3}$
\begin{eqnarray}\label{eq:m2}
\text{Model 2~:}\qquad J^{\text{NP}}(b_T)=e^{-a_1 b_T}\frac{1+a_2 b^2}{1+a_3 b^2}\,,
\end{eqnarray}
with $a_1>0$ and $a_3>0$.

The model 1 is similar to the model used in ref.~\cite{Kang:2024dja}, and is known to accurately describe the data for EEC. The model 2 is used to explore the possible model bias, since the model 1 depends only on a single parameter. It is important to remark that given the value of $B^{*}_J=0.2$ GeV$^{-1}$, the jet function is dominated by its nonperturbative part. Note, that the exponential behavior of this model is also a sign of the severity of the nonperturbative corrections to the jet function, because all $b_T$-corrections come in even powers. We have checked that the data can be described well also by even function, but including very large values of parameters (required to simulate exponential behavior). 

The summary of the theoretical elements that enter our expression is presented in the table \ref{tab:theory}. In general it incorporates many perturbative elements, at N$^3$LO and N$^4$LO. In total it can be characterized as N$^4$LL perturbative accuracy, following the definitions used in the TMD community \cite{Bacchetta:2022awv}.

\begin{table}[h]
\begin{center}
\begin{tabular}{||l|c|l|c||}
\hline
Element & Symbol & Comment & Reference \\
\Xhline{5\arrayrulewidth}
\multicolumn{3}{||c|}{Hard part} & 
\\
\hline
Hard coef.function& $H_{f\bar f}$ & N$^4$LO ($a_s^4$) & \cite{Lee:2022nhh} 
\\
\hline
Cusp AD& $\Gamma_{\text{cusp}}$ & N$^4$LO ($a_s^5$) & \cite{Moch:2018wjh, Herzog:2018kwj} 
\\
\hline
TMD AD& $\gamma_V$ & N$^3$LO ($a_s^4$) & \cite{Lee:2022nhh} 
\\
\hline
Special $\zeta$-line& $\zeta_\mu(b)$ & Exact solution at $a_s^4$  & \cite{Vladimirov:2019bfa, Scimemi:2019cmh}
\\
\Xhline{5\arrayrulewidth}
\multicolumn{3}{||c|}{Collins-Soper kernel} & (\ref{def:CS-kernel})
\\
\hline
Pert.part & $\mathcal{D}_{\text{pert}}$ & N$^3$LO ($a_s^4$) & \cite{Moult:2022xzt, Duhr:2022yyp}
\\
\hline
NP part & $\mathcal{D}_{\text{NP}}$ & 2 parameters $\{c_0, c_1\}$ & (\ref{CS:NP-part})
\\
\hline
Defining scale & $\mu^*$ & & (\ref{CS:scale})
\\
\hline
$b$-prescription & $b^*$ & & (\ref{CS:scale})
\\
\Xhline{5\arrayrulewidth}
\multicolumn{3}{||c|}{Jet function $J$} & (\ref{def:J-f})
\\
\hline
Matching coef. & $\mathbb{J}$ & N$^3$LO ($a_s^3$) & (\ref{def:coef-perturbative})
\\ 
\hline
NP part & $J_{\text{NP}}^f$ & model 1: 1 parameter & (\ref{eq:m2})
\\
 &  & model 2: 3 parameters & (\ref{eq:m2})
\\
\hline
OPE scale & $\mu_{\text{OPE}}$ & & (\ref{mu:OPE})
\\
\Xhline{5\arrayrulewidth}
\multicolumn{3}{||c|}{QCD coupling constant} & 
\\
\hline
Order &  N$^4$LL & 5-loop $\beta(a_s)$ &\cite{Herzog:2017ohr}
\\
\hline
Thresholds &  N$^4$LO ($a_s^4$)& $m_c^{\rm pole},m_b^{\rm pole},m_t^{\rm pole}=1.275,4.18, 172.9$GeV &\cite{Schroder:2005hy}
\\
\hline
\end{tabular}
\end{center}
\caption{\label{tab:theory} Summary of elements used for the theoretical setup. The perturbative order is given with respect to the leading term of the expressions (i.e. leading term is LO). To avoid confusion, in parentheses we designate the power of $\alpha_s$ for the last included perturbative term.}
\end{table}

\section{Review of the data}
\label{sec:data}

\begin{table}[t]
\centering
\begin{tabular}{|c|c|c|}
\hline
Experiment & $Q$ (GeV)& Number of points
\\ \hline
OPAL~\cite{OPAL:1990reb}      &91.0& 8 \\ 
OPAL~\cite{OPAL:1991uui}      &91.3& 30 \\ 
OPAL~\cite{OPAL:1993pnw}      &90.7-91.7  & 17 
\\ \hline 
SLD~\cite{SLD:1994idb} &91.2&8 
\\\hline
TOPAZ~\cite{TOPAZ:1989yod} &53.3 & 8\\ 
       & 59.5& 8
\\ \hline
MARKII~\cite{Wood:1987uf} & 29 &8
\\\hline
TASSO~\cite{TASSO:1987mcs} & 14 &8  \\ 
      & 22 & 8\\ 
      & 34.8 & 8\\ 
      & 43.5&8
\\ \hline
MAC~\cite{Fernandez:1984db} & 29 &8 
\\ \hline
PLUTO ~\cite{PLUTO:1981gcc} & 7.7 &2 \\ 
   & 9.4 &3 \\
    & 12 & 3\\
    & 13 & 3\\
    & 17 & 3\\
    & 22 & 3\\
    & 27.6 & 3\\
    & 27.6-31.6 &40 
  \\ \hline
CELLO ~\cite{CELLO:1982rca} & 22 &8 \\ 
     & 34 & 8
\\ \hline
JADE ~\cite{JADE:1984taa} & 14 & 8\\ 
    & 22 &8 \\ 
    & 34 &8 
\\ \hline
   DELPHI~\cite{DELPHI:1992qrr}&91.2&8\\
   DELPHI~\cite{DELPHI:1993nlw}&91.2&8
\\\hline
Total & & 243\\  \hline
\end{tabular}
\caption{Experiments that have provided data on EEC used in this work, center of mass energy and number of points used in the fit of the present work.}
\label{tab:data-u}
\end{table}

There is a large amount of measurements of EEC. Most part of these data are sufficiently well documented to be used. The synopsis of these data is presented in  table \ref{tab:data-u}. In total, there are 10 experiments which span  $Q$ from 7.7 to 91.7 GeV. All measurements are given as $\chi$ distributions (the second case of eq.\eqref{eq:EECdists}) except for TASSO and the PLUTO $Q=27.6-31.6$GeV set, given as distributions in $\cos\chi$ (the third case of eq.\eqref{eq:EECdists}).  

\noindent Since many of these experiments are old, they require some degree of interpretation. In particular, we could not find any useful information about correlations between errors and therefore we treated all given uncertainties as uncorrelated.  
In the following we summarize our observations related to  each measurement included in the fit. 

\begin{itemize}
\item \textbf{OPAL} We found three separate OPAL publications concerning EEC data. We could not distinguish if these publications present independent measurements or different analysis of the same data. So, we consider these data as independent data sets: 
\begin{itemize}
    \item Ref.\cite{OPAL:1990reb} The measurements of EEC at the center of mass energy of 91GeV\footnote{At the end of section 2, it is mentioned instead the interval $Q=88.3-95.0$GeV for the center of mass energies of the data sample used in the analysis of the publication. The value 91GeV is provided in section 3 and in table 1 for the center of mass energy of the EEC distribution. We used this as exact because, it does not correspond to the central value of the interval and we could not get the corresponding weights to average the result.} are summarized in table 1, including statistical and systematic errors for each bin. Despite the caption of the table indicates correlations between systematic errors of the different bins, they do not provide information about these in the paper nor in HEPData \cite{hepdata.29525}. As explained in section 3 of the corresponding paper, the statistical error was obtained as the root mean square deviation in each bin from 10 different samples of Monte Carlo events and for the systematic uncertainty, they summed in quadrature the values from the different sources. The bin size is 3.6 degrees. 
    \item Ref.~\cite{OPAL:1991uui}  Table 1 shows the values for EEC at 91.3GeV. Althought the center of mass energy is averaged, as mentioned in section 2 of the reference,  the full range is not specified such that we used it as exact. The errors were computed in analogous way as in the previous publication (see section 3 of ref.~\cite{OPAL:1991uui}), but only the combination of statistical and systematic uncertainties are provided. The data, which can also be found in HEPData, ref.~\cite{hepdata.29245}, are binned in 1 degree intervals.      
    \item Ref.~\cite{OPAL:1993pnw} The publication mentions (at the end of sec.~3.2 in ref.~\cite{OPAL:1993pnw}), the use of a covariance matrix to describe the correlations of the bin errors. However, we could not find the values of this matrix and the only uncertainties available are provided in Table 1. They are based on the diagonal terms and include, as specified in the caption, statistical and experimental systematic uncertainties added in quadrature. As appear in sec.~2 of ref.~\cite{OPAL:1993pnw}, the energy considered was within 0.5 GeV of the Z boson mass. So, in order to compare with these data, we averaged formula \eqref{eq:Xsec1} by carrying the integral in $Q$ within the interval and dividing the result by the corresponding width. Same data can be found \footnote{The head of the table contain a typo, specifying the value of center of mass energy as $\sqrt{s}=0.215-0.216$GeV.} also in HEPData, ref.~\cite{hepdata.14427}. The bin size is 1.8 degrees. 
\end{itemize}  
\item \textbf{SLD}~\cite{SLD:1994idb}
Hadron level EEC data can be found in table VI of the paper and also in \mbox{HEPData}~\cite{hepdata.22450}. It is binned in 3.6 degrees intervals and includes statistical and systematic errors obtained as explained in section 5 of the text. 
\item \textbf{TOPAZ}~\cite{TOPAZ:1989yod} As indicated in the reference, events with $Q=52-55$GeV and $Q=59-60$GeV were considered but treated correspondingly as $Q=53.3$ and $Q=59.5$, so we also kept the center of mass energy fixed. The data employed now has bin size of 3.6 degrees and is in tables 1 and 2 but adding the 4$\%$ systematic error specified before figure 1 in the paper. The same appear in HEPData~\cite{hepdata.29800}.
\item \textbf{MARKII}~\cite{Wood:1987uf} EEC measurements at $Q=29$GeV are provided in table I for two different detector configurations, before and after an upgrade. For this work we used the combination of both sets, available at HEPData~\cite{hepdata.23323}. Results were given for 3.6-degree size bins including statistical and systematic errors. 
\item \textbf{TASSO} \cite{TASSO:1987mcs} The publication provides the energy ranges taken into account for each data sample and the corresponding mean values, see table I. We carried out our analysis employing these mean values. Tables II and III contain the EEC distributions binned respectively in $\cos\chi$ and $\chi$. For the analyses we use the data binned in $\chi$ (table III), since it has narrower bins (3.6 degrees). To our surprise, the type of uncertainties given in the tables is not mentioned anywhere in the text. These uncertainties are much smaller than the $10\%$ uncertainty announced in the last sentence of section II as upper limit. Additionally, the procedure to estimate this boundary, detailed in the same section, is the usual one for the systematic uncertainties. All in all, we consider the uncertainties provided in the paper tables (same as on HEPData ~\cite{hepdata.1698}) to be only statistical and add a $10\%$ of systematic uncertainty.  

\item \textbf{MAC}~\cite{Fernandez:1984db} 3.6-degrees binned EEC data with combined statistical and systematic uncertainty is contained in Table I from the reference and also within HEPData~\cite{hepdata.23586}.

\item \textbf{PLUTO}~\cite{PLUTO:1981gcc} The values of EEC for energy values $Q=7.7,\,9.4,\,12,\,13,\,17,\,22,\,27.6,\,30-31.6$GeV are available in the Table I in the HEPData~\cite{hepdata.6228}, in non-uniform bins\footnote{In addition, the angles closer to the back-to-back limit are grouped differently for the lowest energy.} (6 or 12 degrees) in $\chi$. The table II presents the combination of measurements with $Q=27.6-31.6$GeV close to the back-to-back limit using narrow bins ($\Delta\cos\chi=0.001$). In our analyses we use these data instead the data at $Q=30-31.6$GeV from table I. This way we can incorporate more points to the analysis without double counting the data. Since the values of mean $Q$ are not presented, we computed the $Q$-average of the formula \eqref{eq:Xsec1} in the corresponding interval.

The original publication offers  little information regarding uncertainties. Some additional information, can be found in the later paper~\cite{PLUTO:1985yzc} (discussed also below). On the other hand, the header of the HEPData tables indicates that systematic uncertainty were estimated to be smaller than statistical uncertainty. Therefore, we assumed the uncertainties provided were just statistical and we add an estimated systematic uncertainty with the same value as the statistical one.

\item \textbf{CELLO}~\cite{CELLO:1982rca} The measurements obtained for the energy correlator are just on HEPData, ref.~\cite{hepdata.16413}, where one error for each 3.6-degrees bin is shown. Since we can describe these data and we did not find any comments about the EEC uncertainties we consider them total errors from the combination of statistical and systematic ones. 

\item \textbf{JADE}~\cite{JADE:1984taa} As explained in the abstract and data section of the corresponding publication, the experiment had energies fixed at $Q=14$GeV and $Q=22$GeV but also events with $Q=29-36$GeV were combined together in a third set. For this last case, the average energy value $<Q>\,=34$GeV is provided such that one can see that the events were not distributed uniformly within the interval and therefore we decided to carry out our analysis using this fixed quantity. 

The values for EEC are given with 3.6-degrees size $\chi$-intervals in table 1 and in HEPData, ref.~\cite{hepdata.1998}. As in the previous case, only one type of uncertainty is present. We assume that the reported uncertainty accounts for statistical and systematic contributions, given that no further information was found.

\item \textbf{DELPHI} To our knowledge, three different sets of data from this collaboration are available. The following are the ones we include in our study:
\begin{itemize}
\item Ref.\cite{DELPHI:1992qrr}  As mentioned in the introduction, EEC was obtained from the sample of events collected in 1990, and presented in table 1f. This table, with 3.6-degree bins and the two types of uncertainties, also appears in HEPData, ref.~\cite{hepdata.14603}.
\item Ref.\cite{DELPHI:1993nlw} Here, the events collected in 1991 were used.  Results are given in table 6 and in HEPData, ref.~\cite{hepdata.50115}  with the same bin width as before, also including statistical and systematic uncertainties but only for $\chi\geq 90^{\degree}$.  
\end{itemize}
\end{itemize}

Finally, most of the data sets (apart of the sets that do not cover the whole range of $\chi$) are normalized to 1 in the interval $\chi\in [0^{\degree}, 180^{\degree}]$. With a unique exception of PLUTO, which is normalized to $2$.  Note, that TMD analyses does not provide the normalization and that is why, within our approach, we cannot describe the normalization for all cases as it is discussed in section \ref{sec:stat}.

\begin{figure}[t]
\centering
\includegraphics[width=0.5\linewidth]{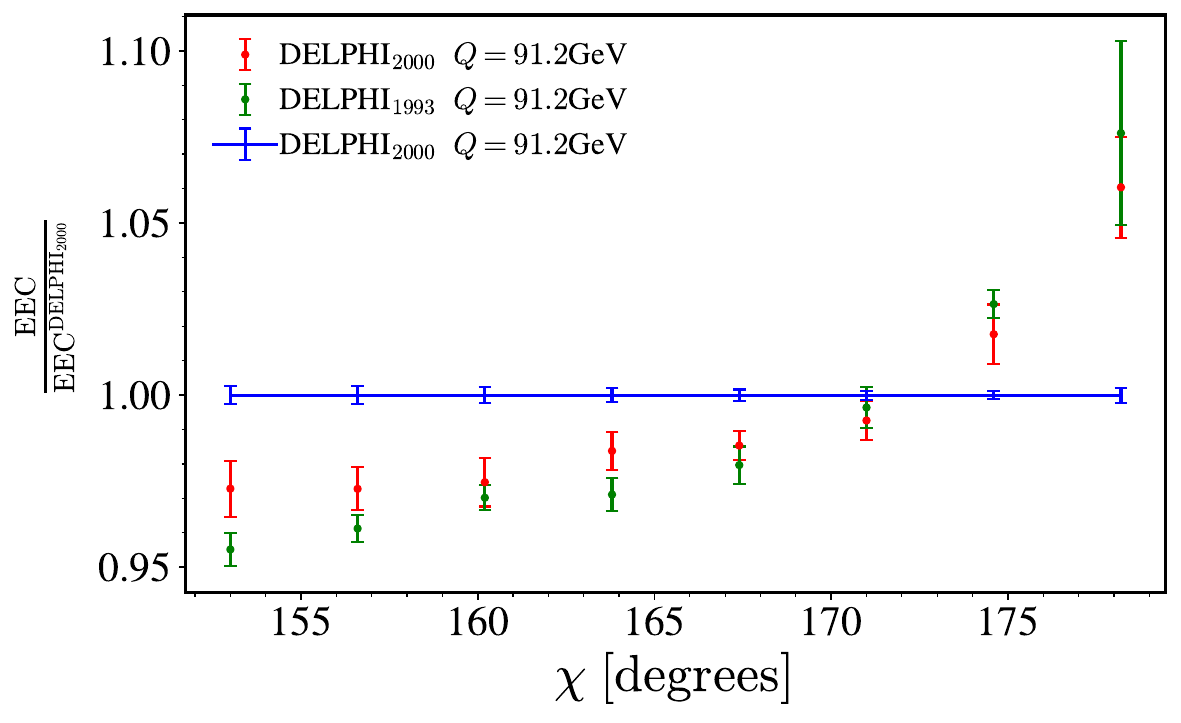}
\caption{Comparison between whole $\theta_T$-range data from \cite{DELPHI:2000uri}: DELPHI$_{2000}$ with \cite{DELPHI:1992qrr}: DELPHI$_{1992}$   and \cite{DELPHI:1993nlw}: DELPHI$_{1993}$. The three sets are shown normalized to one within the back-to-back region}
\label{fig:DELPHI}
\end{figure}

Furthermore, we have found significant discrepancy between our predictions and a few data sets. For that reason, we exclude these data sets from our analyses. Let us briefly comment on these measurements, and the reason of their exclusion:
\begin{itemize}
\item \textbf{SLD}~\cite{SLD:1994yoe} This is an early measurement by SLD collaboration that is later superseded\footnote{Ref.~\cite{SLD:1994yoe} used events recorded in 1992, as mentioned in section II of the corresponding paper, whereas in Ref.~\cite{SLD:1994idb} 1992 and 1993 data samples are combined, as specified in its introduction.} by \cite{SLD:1994idb} (included in our fit). These data is only available via HEPData, ref.~\cite{hepdata.17744}, and it contains an unknown factor that corrects for unspecified hadronization effects.
\item \textbf{PLUTO}~\cite{PLUTO:1985yzc} These data taken at $Q=34.6$ has extremely small uncertainties (of order of $1\%$) compared to their previous publication. In the reference, the improvement on the statistics is remarked while the systematic uncertainty for EEC is not documented. Although, our prediction for these data is in a general agreement, we were not able to describe these data statistically-well within the announced precision of points.
\item \textbf{DELPHI} ~\cite{DELPHI:2000uri} This EEC distribution is given in HEPData, ref.~\cite{hepdata.13245}, for different values of the polar angle of thrust axis in order to improve the precision by taking into account the dependence of the detector properties with respect to this angle. In the introduction of the paper, it is pointed out that smaller uncertainties were obtained in comparison to refs.~\cite{DELPHI:1993nlw,DELPHI:1992qrr}. However, the direct comparison of \cite{DELPHI:1993nlw, DELPHI:1992qrr}  and ref.\cite{DELPHI:2000uri} shows discrepancy between them, see fig.\ref{fig:DELPHI}. We have not found any discussion of this discrepancy in the publication.

\item \textbf{ALEPH}~\cite{ALEPH:1990vew} The data tables are lost.
\end{itemize}

\begin{figure}[bt]
\centering
\includegraphics[width=0.5\linewidth]{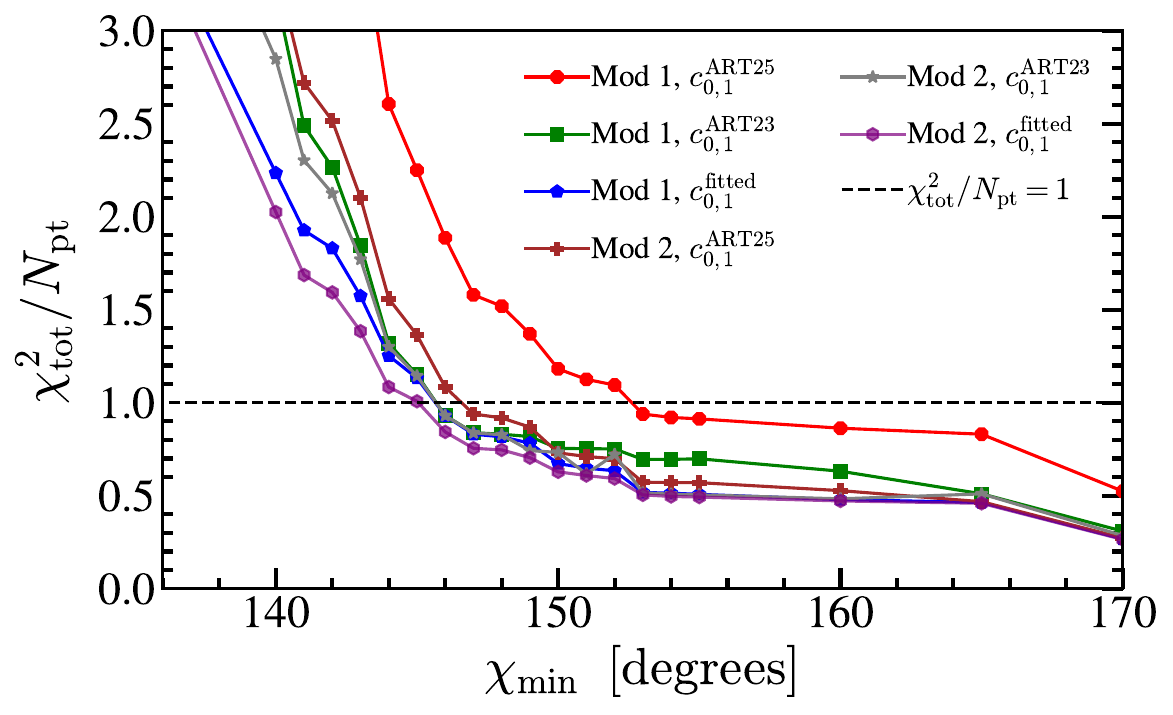}
\caption{Test of $\chi^2/N_{\text{points}}$ as a function of the parameter $\chi$.}
\label{fig:chicut}
\end{figure}

The TMD factorization is valid in the limit $\chi \to 180^o$. Practically, it implies that one should introduce a cut value $\chi_{\text{min}}$, below which TMD factorization is assumed to be violated. The data points with $\chi<\chi_{\text{min}}$ are to be excluded from the fits. Since there is no physical scale to fix this parameter, it should be identified by other means. To identify $\chi_{\text{min}}$ we used the same strategy as the one utilized in the other TMD studies \cite{Scimemi:2017etj, Scimemi:2019cmh, Bacchetta:2019sam, Bacchetta:2022awv, Moos:2023yfa, Moos:2025sal}. Namely, we investigate the convergence of the fit with respect to different values of $\chi_{\text{min}}$. Starting from the large values of $\chi_{\text{min}}$ (where TMD factorization is definitely valid) we decrease it, including more points, and repeating the fit. At a certain moment, the parameters of TMD factorization are not capable to incorporate effects of growing power corrections, and the value of the likelihood function ($\chi^2$-test function defined below) drastically increases. The change of behavior identifies the value of $\chi_{\text{min}}$. 

In fig.~\ref{fig:chicut} we plot the values of $\chi^2/N_{\text{pt}}$ obtained in different models at different $\chi_{\text{min}}$. One can clearly see that the change of behavior happens at $\chi \simeq150^{o}$. This value corresponds to an effective $q_T^{\text{eff}}=\sqrt{1-z}Q\approx 0.26Q$. This is in perfect agreement with the estimations obtained from SIDIS and Drell-Yan data, where it was found that TMD factorization works up to $q_T\simeq 0.25 Q$.

So, we select only points with $\chi\geqslant\chi_{\text{min}}=150^{o}$, that corresponds to 243 points in total. Their distribution with respect to experiments is shown in table \ref{tab:data-u}.

\section{Statistical analysis and  theory normalization}
\label{sec:stat}

The fitting procedure and the method of estimation of uncertainties are identical to the ones employed in ART23 and ART25 \cite{Moos:2023yfa, Moos:2025sal}, which in turn are adopted from NNPDF collaboration ~\cite{Ball:2008by, Ball:2012wy}. For completeness, we present the main elements here, referring the reader to the mentioned articles for any missing detail.

We use the following definition of the $\chi^2$-test function
\begin{eqnarray}\label{def:chi2-test}
\chi^2=\sum_{i,j\in \text{data}}(m_i-t_i)V^{-1}_{ij}(m_j-t_j),
\end{eqnarray}
where $i$ and $j$ run over all data points included in the fit, $m_i$ and $t_i$ are the experimental value and theoretical prediction for point $i$, respectively, and $V^{-1}_{ij}$ is the inverse of the covariance matrix. The covariance matrix is defined as
\begin{eqnarray}\label{def:V}
V_{ij}=\delta_{ij}V_i^{\text{uncorr.}}+V_{ij}^{\text{corr.}}=\delta_{ij}\Delta_{i,\text{uncorr}.}^2+\sum_{l}\Delta_{i,\text{corr}.}^{(l)}\Delta_{j,\text{corr}.}^{(l)},
\end{eqnarray}
where $\Delta_{i,\text{uncorr}.}$ is the uncorrelated uncertainty of measurement $i$, and $\Delta_{i,\text{corr}.}^{(l)}$ is the $l$-th correlated uncertainty. If the data have a normalization uncertainty (usually due to the uncertainty in the measured luminosity), it is included in the $\chi^2$ as one of the correlated uncertainties.

Along the analyses of the EEC data, we have observed that many data sets agree with our prediction by shape, but disagree in the normalization. The normalization disagreement could be small in absolute values, but it significantly spoils the agreement between theory and data in statistical terms. Practically, it inflates the $\chi^2/N$ by factor 2-3.

Given the  oldness of data it is not possible to reconstruct the origin of the normalization discrepancy. It may be rooted in the incompleteness of our theoretical normalization factor eq.~(\ref{normalization-factor}), or have experimental origin. In fact, many data sets were normalized to the cross section integrated on all values of $\chi$, which involves utmost regions that where not covered by detector and reconstructed by some other means. Furthermore, experiment normalizes EEC to the detected cross-section, while theory account for all produced particles including ultra-soft ones. Uncertainties due to these assumptions should be reported as additional correlated uncertainty, but they are not.

Thus, in order to be able to statistically describe the data, which  are reported with very small  uncertainties, we normalize our prediction to the data. For it, we use the ``best $\chi^2$''-strategy. We present the theory prediction in the form
\begin{equation}
t_i=\frac{\bar t_i}{1+\delta n},
\end{equation}
where $\delta n$ represent the normalization disagreement. The number $\delta n$ is common for each data set. The value of $\delta n$ is determined by the minimum of $\chi^2$, eq.~ (\ref{def:chi2-test}), and reads
\begin{eqnarray}
\delta n=\sum_{i,j\in \text{data}}(t_i-m_i)V^{-1}_{ij}t_j\Bigg/\sum_{i,j\in \text{data}}m_iV^{-1}_{ij}t_j.
\end{eqnarray}
In the final table we report the values of $\delta n$ for each experiment. Let us remark that for the majority of experiments, we found $\delta n<$2\%, and only for few of them $\delta n\sim 10$\%.

\begin{figure}[t]
\centering
\includegraphics[width=0.5\textwidth]{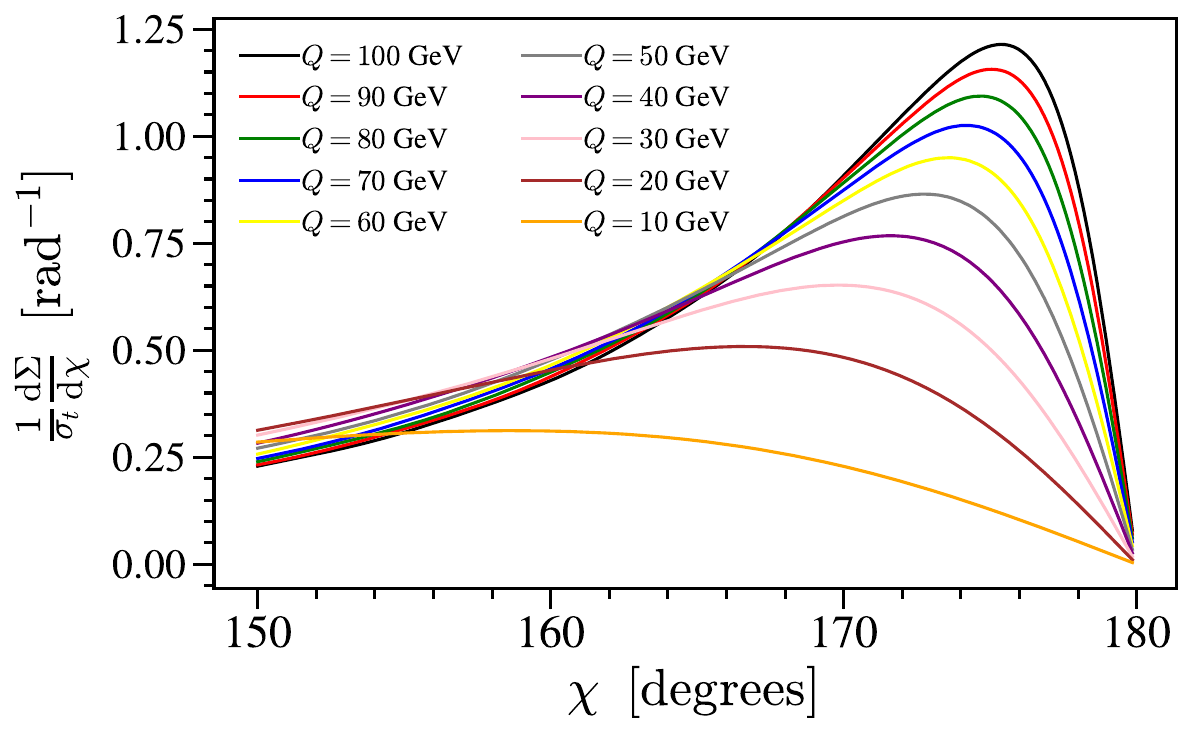}
\caption{Demonstration of the shape-change for EEC with the increment of the energy.}
\label{fig:Q-dependance}
\end{figure}

Normalizing to the data we certainly miss a part of information about the NP parameters. However, it is not decisive, because  most of the restrictions come from the shape of EEC, rather than from the absolute value. This can be seen in fig.~\ref{fig:Q-dependance}, where we show the variation of shape due to the changing of $Q$.

The uncertainty for the parameters of the fit are determined using the Monte-Carlo replica method. It consists in the resampling of the data according to its uncertainty and finding the minimum of $\chi^2$-test function for this re-sampled data. This process produces an ensemble of parameters which serves to determine mean value and uncertainties of any secondary observable. For the details of this procedure see refs.~\cite{Ball:2008by, Ball:2012wy}. 

The replica method is known for its robustness and simplicity of implementation, and thus, is used by many phenomenological groups, including ref.~\cite{Kang:2024dja}. However, as we demonstrate in the following section, this method is not applicable to the present data set, or better to say, requires critical re-examination of the data and/or definition of the test function.

\section{Fit results and issues of the uncertainty determination}
\label{sec:fit}

\begin{table}
\begin{center}
\begin{tabular}{|c|c!{\vrule width 2pt}c|c!{\vrule width 2pt}c|c!{\vrule width 2pt}c|c|}
\hline
Experiment  &Q&\multicolumn{2}{|c!{\vrule width 2pt}}{$c_{0,1}^{\text{ART25}}$}&\multicolumn{2}{|c!{\vrule width 2pt}}{$c_{0,1}^{\text{ART23}}$}&\multicolumn{2}{|c|}{$c_{0,1}^{\text{fitted}}$}
\\    
($N_{pt}$)&(GeV)& $\chi^2/N_{pt}$ &$\delta n$ $\%$ & $\chi^2/N_{pt}$ & $\delta n$ $\%$ & $\chi^2/N_{pt}$ & $\delta n$ $\%$ 
\\ \hline   
OPAL~\cite{OPAL:1990reb}      (8)&91.0&0.25&2.09&0.12&2.48&0.14&2.28\\
OPAL~\cite{OPAL:1991uui}      (30)&91.3&1.37&10.3&0.68&12.2&0.54&11.2\\ 
OPAL~\cite{OPAL:1993pnw}      (17)&90.7-91.7  &  1.33&9.96&2.17&10.9&1.57&10.3
\\ \hline 
SLD~\cite{SLD:1994idb}      (8)&91.2 & 0.5&7.44&0.52&9.25&0.3&8.38
\\ \hline
TOPAZ~\cite{TOPAZ:1989yod} (16) &53.3-59.5& 1.87&3.23&1&3.68&1.21&3.46
\\ \hline
MARKII~\cite{Wood:1987uf} (8) &29& 2.55&14.5&2.17&17.1&1.38&16.1
\\ \hline
TASSO~\cite{TASSO:1987mcs} (8) &14& 2.08&1.98&1.21&2.14&1.14&2.04\\
(8)&22&1.7&1.57&0.91&1.76&0.94&1.68\\
(8)&34.8&1.15&1.31&0.53&1.48&0.61&1.39\\
(8)&43.5&0.414&0.928&0.112&1.07&0.15&1
\\\hline
MAC~\cite{Fernandez:1984db} (8) &29& 1.09&8&1.8&9.16&1.34&8.63
\\ \hline
PLUTO ~\cite{PLUTO:1981gcc}  (60) &7.7-31.6& 0.2&0.04&0.18&0.06&0.19&0.05
\\ \hline
CELLO ~\cite{CELLO:1982rca}  (8) &22& 0.85&2.9&0.31&3.5&0.31&3.28\\ 
(8)&34&2.3&2.18&1.27&2.94&1.34&2.64
\\\hline
JADE ~\cite{JADE:1984taa}  (8)&14 & 2.26&9.95&0.96&11.3&0.92&10.7\\ 
(8)&22&1.17&7.33&0.62&8.79&0.54&8.29\\
(8)&34&3.54&9.75&0.86&11.4&1.19&10.7
\\\hline
DELPHI ~\cite{DELPHI:1992qrr}  (8)&91.2 &1.23&6.3&0.54&7.36&0.48&6.74\\ 
~\cite{DELPHI:1993nlw}   (8)&91.2& 1.63&8&0.46&8.9&0.49&8.4
\\ \Xhline{5\arrayrulewidth}
Total (243)&& 1.18&&0.75&&0.67&
\\  \hline
\end{tabular}
\caption{\label{tab:chi2_Mod1}
Summary of values for $\chi^2$ and $\delta n$ for various experiments and obtained within fits with the model 1. 
}
\vspace{0.5cm}
\def\arraystretch{1.5}

\begin{tabular}{|c|c|c|}
\hline
~~$c_0^{\text{ART25}}=8.59^{+0.23}_{-0.17}\cdot 10^{-2}$~~&
~~$c_0^{\text{ART23}}=3.69^{+0.65}_{-0.61}\cdot 10^{-2}$~~&
~~$c_0=3.17^{+0.04}_{-0.03} \cdot 10^{-2}$~~  
\\ 
$c_1^{\text{ART25}}=3.03^{+0.38}_{-0.41}\cdot 10^{-2}$ & 
$c_1^{\text{ART23}}=5.82^{+0.64}_{-0.88}\cdot 10^{-2}$&
$c_1=2.37^{+0.07}_{-0.1} \cdot 10^{-2}$ 
\\ 
{$a_1=0.879^{+0.014}_{-0.015}$ } & 
$a_1=1.002^{+0.009}_{-0.016}$  &
$a_1=0.941^{+0.002}_{-0.003}$  
\\\hline
\end{tabular}
\caption{Values of parameters obtained within the fits with model 1 $J^{\text{NP}}(b_T)=\exp(-a_1 b_T)$. The parameters are presented in units of GeV, (i.e. $c_{i=0,1}$(GeV$^{-2}$), $a_1$(GeV), $a_{2,3}$(GeV$^2$)).
}
\label{tab:fit_Mod1}
\end{center}
\end{table}

\begin{table}[ht]
\begin{center}
\begin{tabular}{|c|c!{\vrule width 2pt}c|c!{\vrule width 2pt}c|c!{\vrule width 2pt}c|c|}
\hline
Experiment  &Q&\multicolumn{2}{|c!{\vrule width 2pt}}{$c_{0,1}^{\text{ART25}}$}&\multicolumn{2}{|c!{\vrule width 2pt}}{$c_{0,1}^{\text{ART23}}$}&\multicolumn{2}{|c|}{$c_{0,1}^{\text{fitted}}$}\\    
($N_{pt}$)&(GeV)& $\chi^2/N_{pt}$ &$\delta n$ $\%$ & $\chi^2/N_{pt}$ & $\delta n$ $\%$ & $\chi^2/N_{pt}$ & $\delta n$ $\%$ 
\\ \hline   
OPAL~\cite{OPAL:1990reb}      (8)&91.0&0.13&2.26&0.13&2.45&0.16&2.34\\ 
OPAL~\cite{OPAL:1991uui}      (30)&91.3&0.48&11.1&0.57&12.1&0.62&11.6  \\ 
OPAL~\cite{OPAL:1993pnw}      (17)&90.7-91.7  &  1.6&10.2&1.82&10.8&1.36&10.6
\\ \hline 
SLD~\cite{SLD:1994idb}      (8)&91.2 & 0.27&8.27&0.43&9.1&0.33&8.62
\\ \hline
TOPAZ~\cite{TOPAZ:1989yod} (16) &53.3-59.5& 1.45&3.5&1.06&3.64&1.03&3.45
\\ \hline
MARKII~\cite{Wood:1987uf} (8) &29& 2.01&16.1&1.72&16.8&0.77&16.4
\\ \hline
TASSO~\cite{TASSO:1987mcs} (8) &14& 1.02&1.94&1.49&2.16&1.15&2.19\\
(8)&22&0.95&1.66&1.09&1.75&0.94&1.73\\
(8)&34.8&0.71&1.41&0.63&1.46&0.55&1.39\\
(8)&43.5&0.21&1.02&0.15&1.06&0.1&0.99
\\\hline
MAC~\cite{Fernandez:1984db} (8) &29& 1.48&8.55&1.42&9.06&1.05&8.97\\ \hline
PLUTO ~\cite{PLUTO:1981gcc}  (60) &7.7-31.6& 0.19&0.05&0.19&0.06&0.19&0.06
\\ \hline
CELLO ~\cite{CELLO:1982rca}  (8) &22& 0.35&3.23&0.37&3.44&0.32&3.41\\ 
(8)&34&1.54&2.65&1.41&2.85&1.2&2.65
\\\hline
JADE ~\cite{JADE:1984taa}  (8)&14 &1.01&10.2&1.2&11.3&0.88&11.6\\ 
(8)&22&0.75&8.16&0.6&8.63&0.35&8.59\\
(8)&34&1.41&10.8&1.14&11.2&1.29&10.8
\\\hline
DELPHI ~\cite{DELPHI:1992qrr}  (8)&91.2 &0.38&6.72&0.54&7.28&0.81&6.97\\ 
~\cite{DELPHI:1993nlw}   (8)&91.2& 0.45&8.33&0.41&8.82&0.54&8.68
\\ \Xhline{5\arrayrulewidth}
Total (243)&& 0.73&&0.73&&0.63&
\\  \hline  
\end{tabular}
\caption{\label{tab:chi2_Mod2}
Summary of values for $\chi^2$ and $\delta n$ for various experiments and obtained within fits with the model 2. 
}
\vspace{0.5cm}
\def\arraystretch{1.5}
\begin{tabular}{|c|c|c|}
\hline
~~$c_0^{\text{ART25}}=8.59^{+0.23}_{-0.17}\cdot 10^{-2}$~~&
~~$c_0^{\text{ART23}}=3.69^{+0.65}_{-0.61}\cdot 10^{-2}$~~&
~~$c_0=1.41^{+0.22}_{-0.19} \cdot 10^{-2}$~~  
\\ 
$c_1^{\text{ART25}}=3.03^{+0.38}_{-0.41}\cdot 10^{-2}$ & 
$c_1^{\text{ART23}}=5.82^{+0.64}_{-0.88}\cdot 10^{-2}$&
$c_1=5.31^{+0.34}_{-0.23} \cdot 10^{-2}$ 
\\ 
{$a_1=0.976^{+0.013}_{-0.013}$ } & 
$a_1=0.949^{+0.033}_{-0.030}$  &
$a_1=0.917^{+0.003}_{-0.003}$  
\\
{$a_2=0.246^{+0.017}_{-0.018}$ } & 
{$a_2=0.83^{+0.95}_{-0.85}$ } &
{ $a_2=0.88^{+0.04}_{-0.04}$ }
\\
$a_3=(5.^{+10.}_{-4.7})\times10^{-7}$  & 
$a_3=1.0^{+1.1}_{-1.0}$ &
{$a_3=1.2^{+0.1}_{-0.1}$  }
\\
\hline
\end{tabular}
\caption{Values of parameters obtained within the fits with model 2 $J^{\text{NP}}(b_T)=\exp(-a_1 b_T)\frac{1+a_2b_T^2}{1+a_3b_T^2}$. The parameters are presented in units of GeV, (i.e. $c_{i=0,1}$(GeV$^{-2}$), $a_1$(GeV), $a_{2,3}$(GeV$^2$)).
}
\label{tab:fit_Mod2}
\end{center}
\end{table}

We have performed three types of fits:
\begin{enumerate}
\item fitting parameters of the jet-function using CS kernel determined in ART23 eq.~(\ref{CS:ART23}),
\item fitting parameters of the jet-function using CS kernel determined in ART25  eq.~(\ref{CS:ART25}),
\item fitting parameters of the jet-function together with parameters $c_{0,1}$ of the CS-kernel.
\end{enumerate}
Each fit  was done for both models in eq.~(\ref{eq:m1}, \ref{eq:m2}) of jet functions (6 independent fits in total). For each of these fits we made the central fit, and the estimation of uncertainties using the replica method.

The  values of $\chi^2$ and $\delta n$ obtained for these fits are presented in tab.~\ref{tab:chi2_Mod1} and \ref{tab:chi2_Mod2}. Clearly, all models present very good agreement with the data. The worst results are shown by the model 1 with ART25 input, which has $\chi^2/N_{pt}=1.18$. The rest of models have $\chi^2/N_{pt}$ smaller than 1. Moreover, most of experimental sets (13 out of 19) have $\chi^2/N_{pt}<1$. Even if one drops the data sets with $\chi^2/N_{pt}<0.5$ the result is $\chi^2/N_{pt}\sim 0.9$. This behavior is worrisome because it demonstrates that these data are problematic and have hidden correlations.

\begin{figure}[t]
    \centering
	\includegraphics[width=0.85\textwidth]{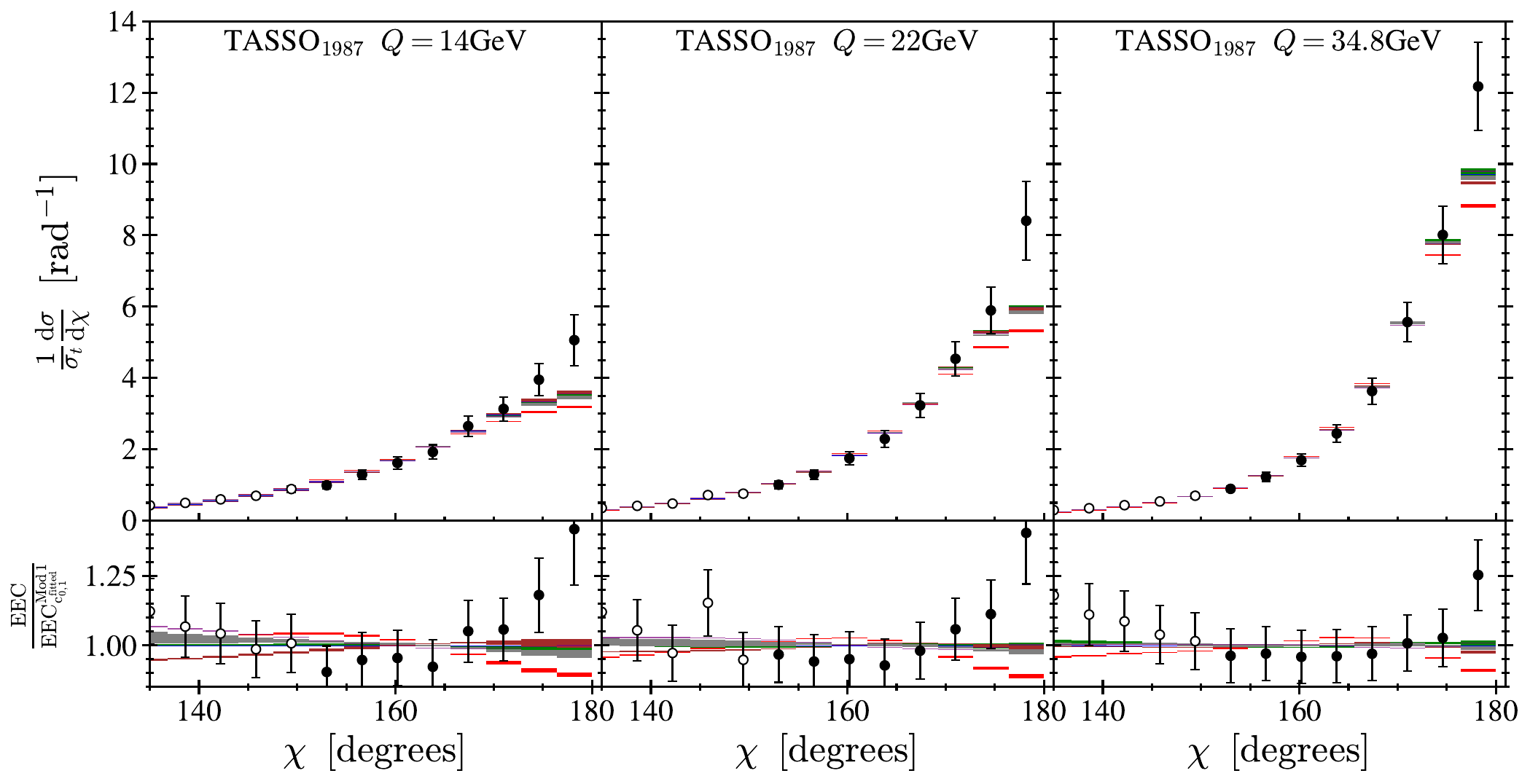}
	\includegraphics[width=0.85\textwidth]{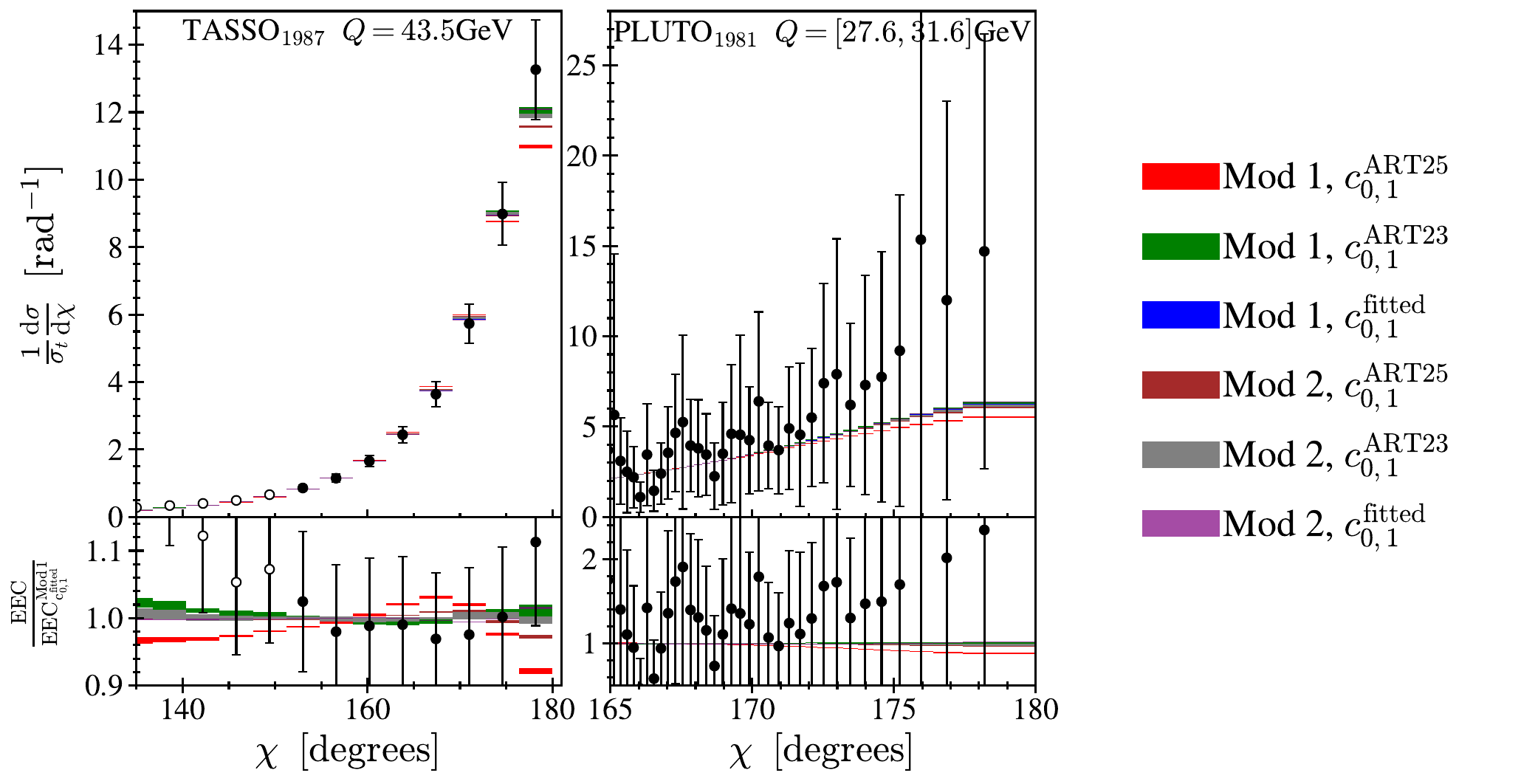}
   \caption{Comparison of the data and the theory prediction for the measurements at low-energy differential in $\cos\chi$. Color of the theory prediction indicates the used model as indicated in the legend. Points indicated by filled(empty) circles were (not) included in the fit.}
\label{fig:fit-low_coschi}
\end{figure}
\begin{figure}[t]
    \centering
	\includegraphics[width=0.85\textwidth]{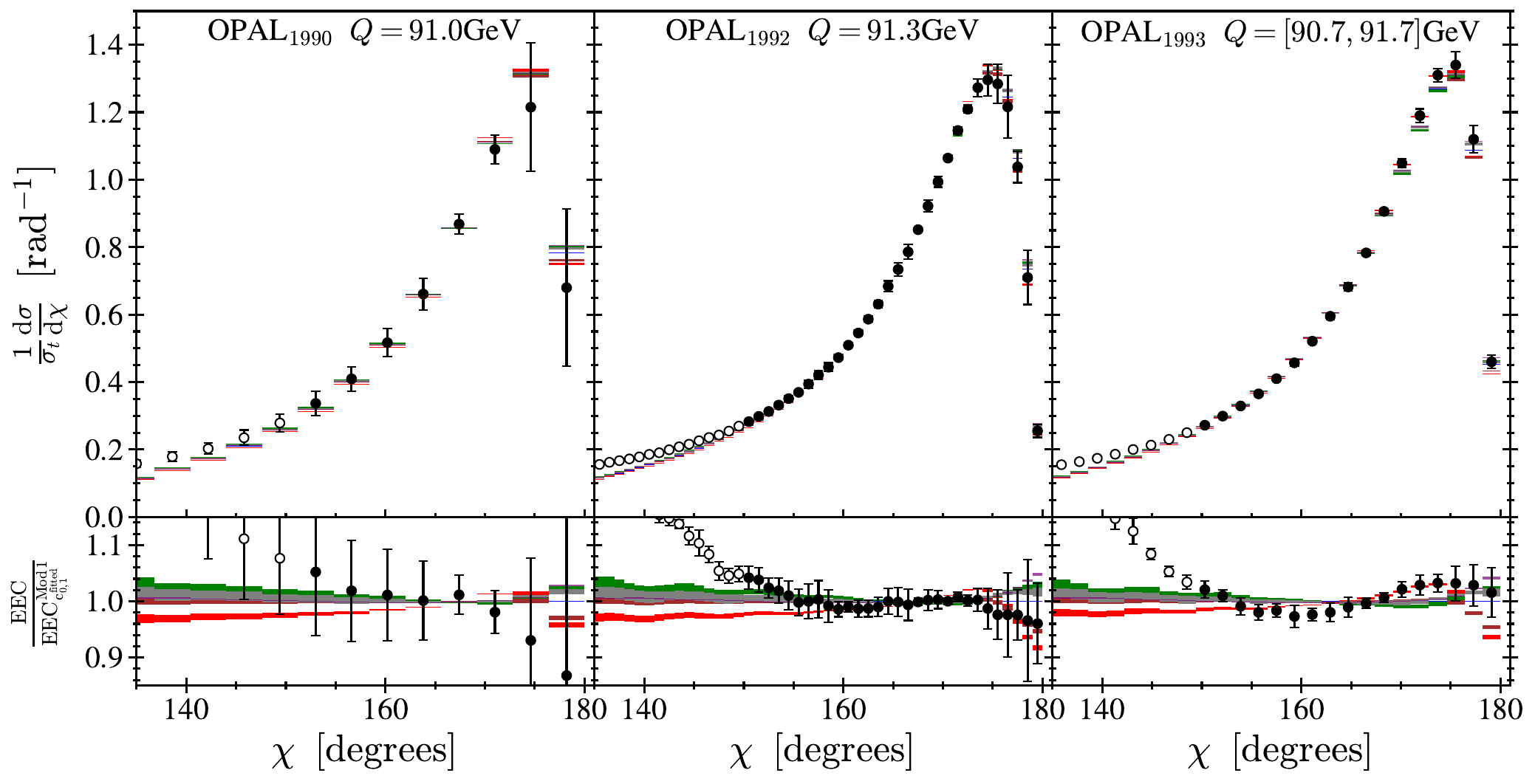}
	\includegraphics[width=0.85\textwidth]{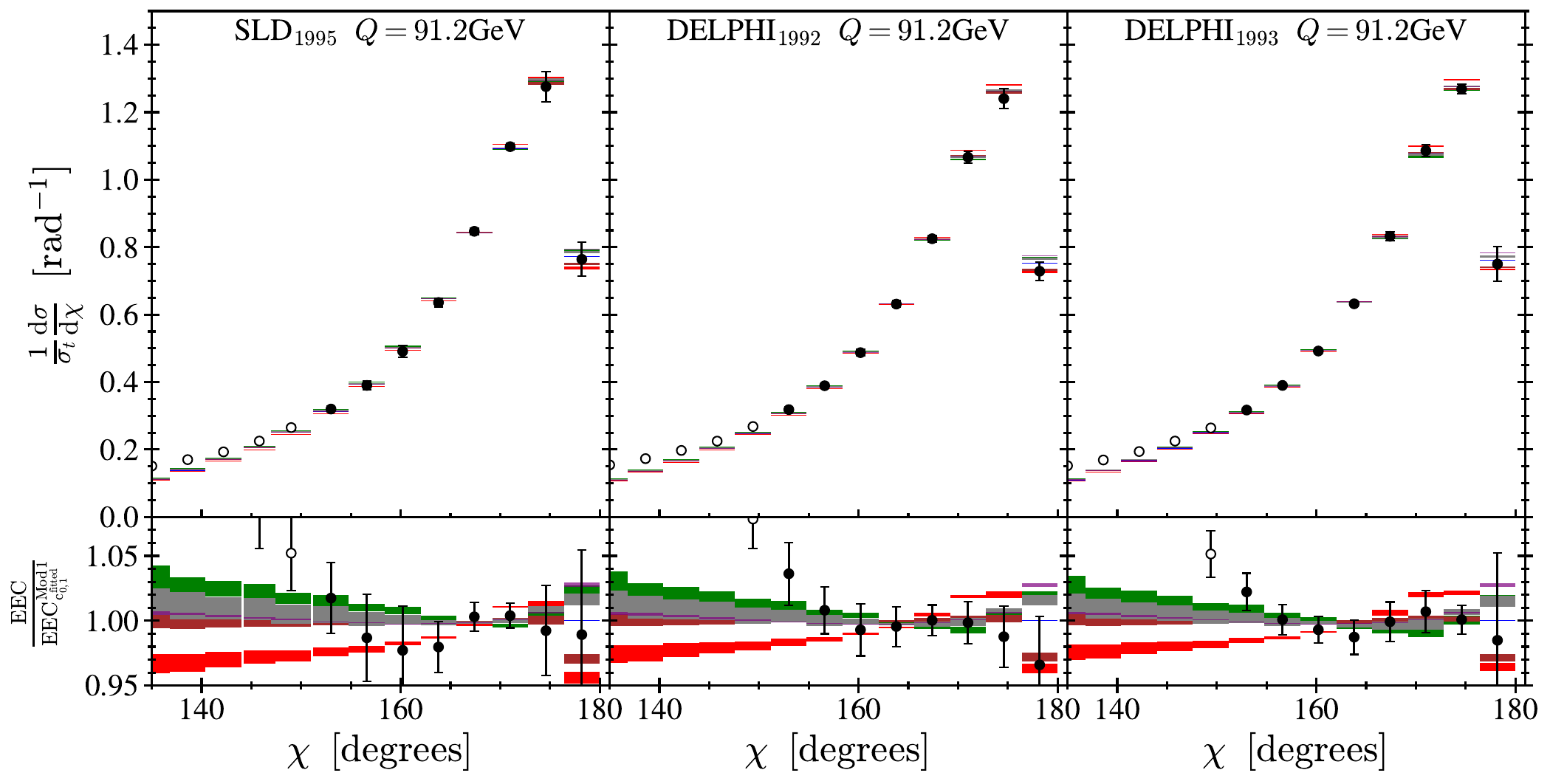}
    \caption{Comparison of the data and the theory prediction for measurements with the energy near the Z-boson mass. Color of the theory prediction indicates the used model as indicated in the legend of the fig.~\ref{fig:fit-low_coschi}. Points indicated by filled(empty) circles were (not) included in the fit.}
\label{fig:fit-Z}
\end{figure}
\begin{figure}[t]
    \centering
	\includegraphics[width=0.85\textwidth]{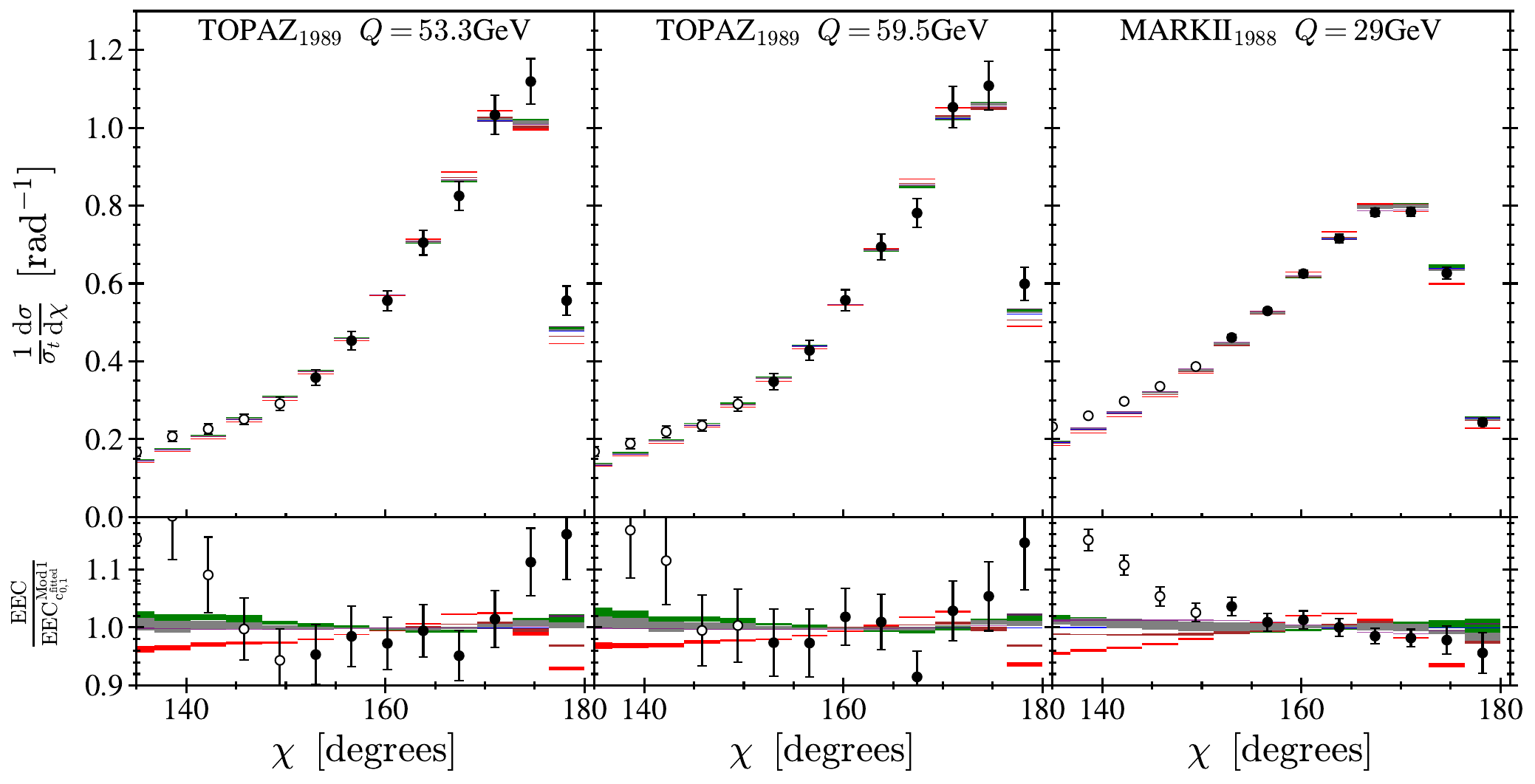}
	\includegraphics[width=0.85\textwidth]{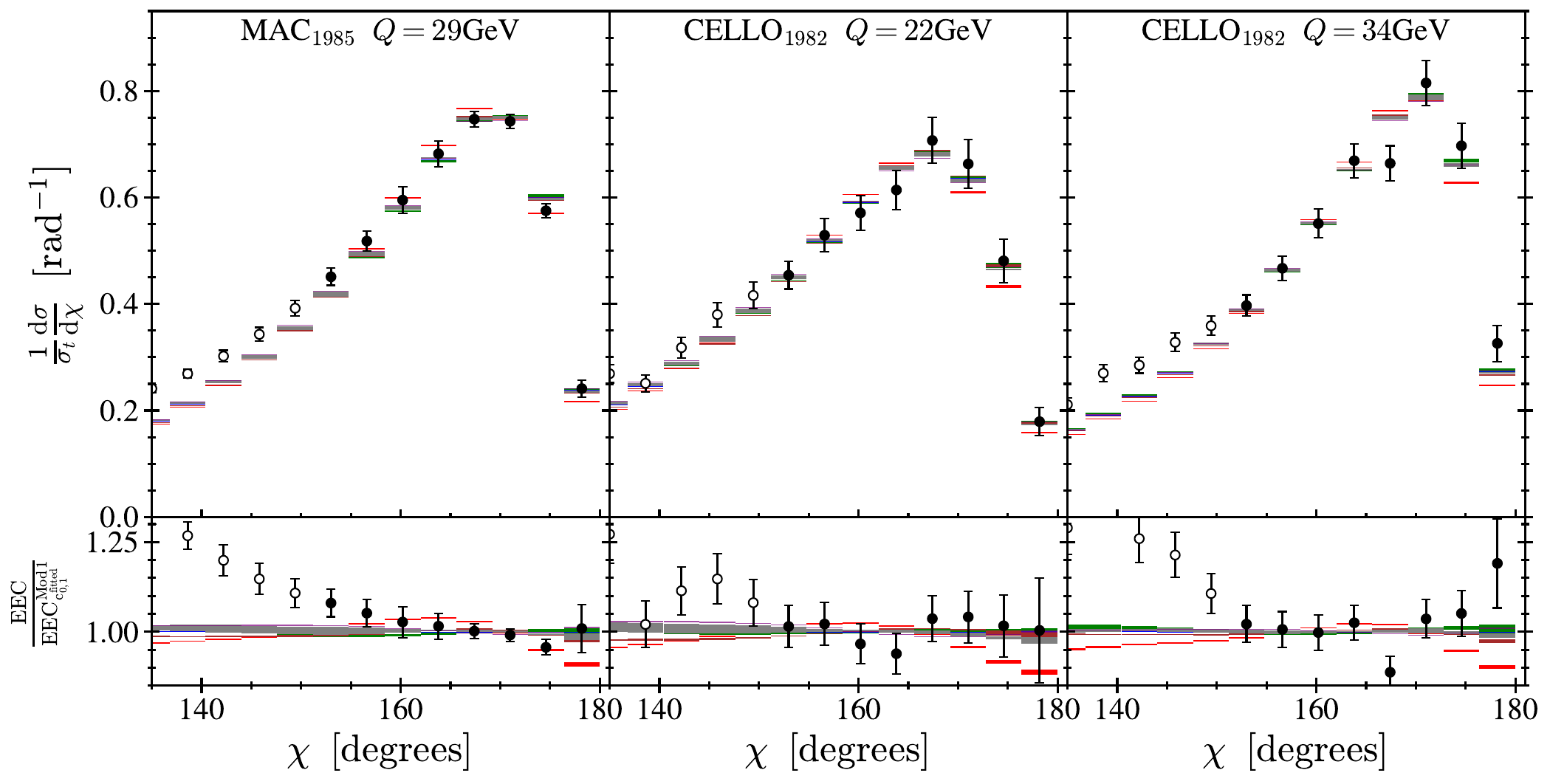}
    \includegraphics[width=0.85\textwidth]{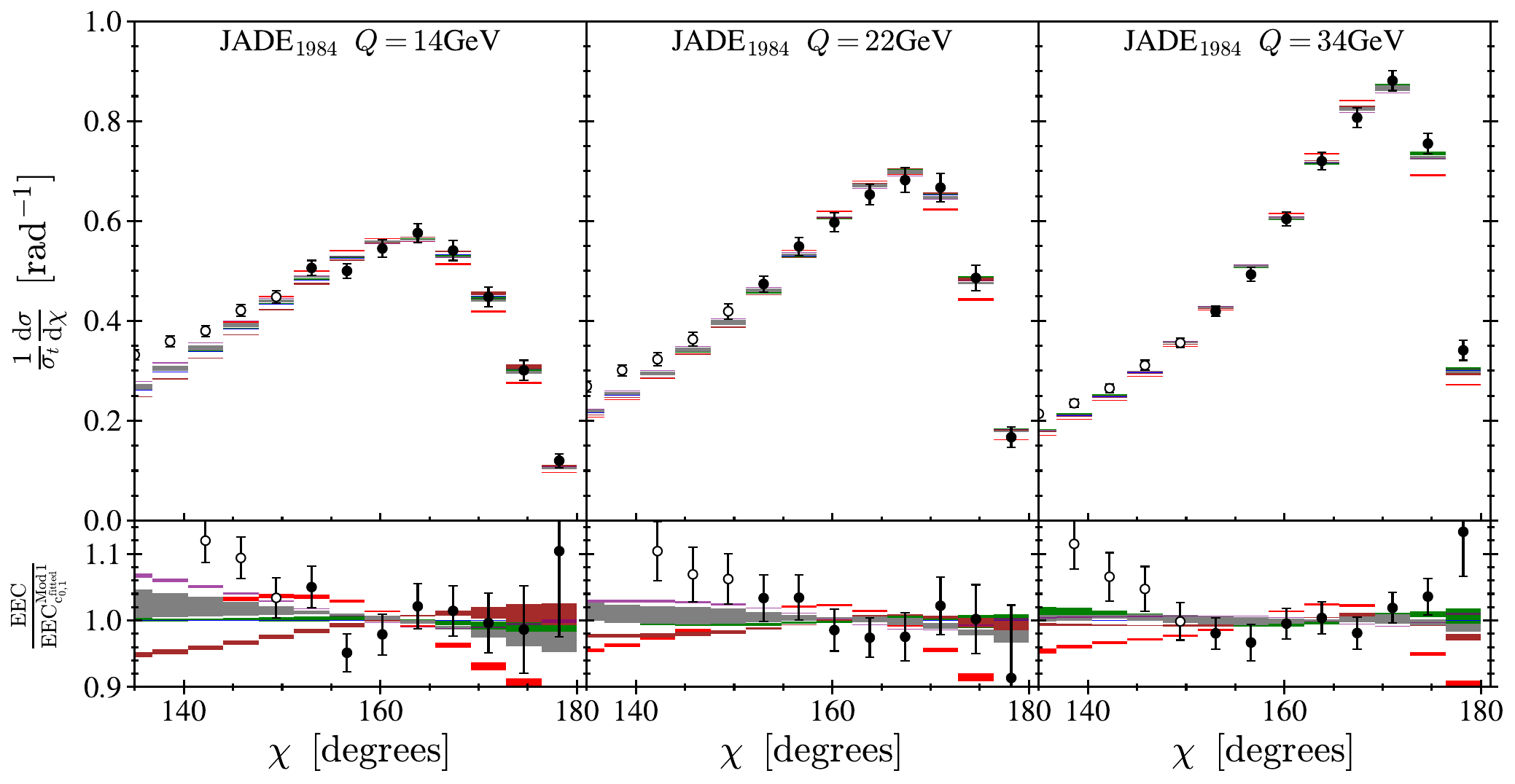}
    \caption{
Comparison of the data and the theory prediction for the measurements at low-energy differential in $\chi$. Color of the theory prediction indicates the used model as indicated in the legend of the fig.~\ref{fig:fit-low_coschi}. Points indicated by filled(empty) circles were (not) included in the fit.}
 \label{fig:fit-low_chi}
\end{figure}

In fig.~\ref{fig:fit-low_coschi}, \ref{fig:fit-Z}, \ref{fig:fit-low_chi}  we demonstrate the comparison of data with the theory prediction for all considered models. All models demonstrate a very similar behavior with minor differences at the $\chi\to180^o$. At $\chi\sim 150^o$ the prediction of the TMD factorization theorem starts to deviate from the measurement. This is due to appearance of the power corrections.

We have not observed any statistically significant improvement between model 1 and model 2. The values of extracted parameters are presented in the tables \ref{tab:fit_Mod1} and \ref{tab:fit_Mod2}. As one can see, the values of jet-function parameters obtained for different CS kernel input are in general agreement. The only counter example are the parameters $a_{2,3}$ in the ART25 setting, which could be explained by the correlation between parameters. 

\begin{figure}
\centering
\includegraphics[width=0.4\textwidth,valign=t]{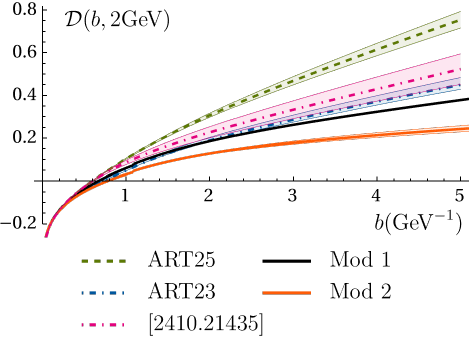}
\includegraphics[width=0.4\textwidth,valign=t]{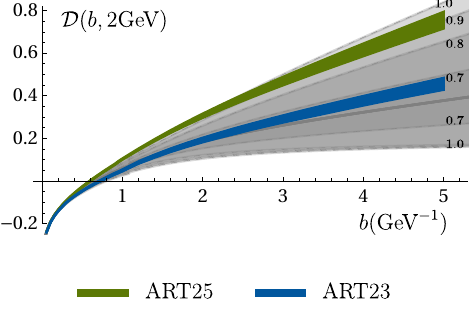}
\caption{\label{fig:CS} (left) Comparison of CS kernels obtained in ART fits, and from the fit of EEC. (right) The ranges of CS kernel at fixed $c_0$ and $c_1$ and fitted parameter $a_1$ (model 1), at different values of $\chi^2/N$ (indicated in the plot).}
\end{figure}

One of the main motivations for this study was to check how well the EEC data can constraint the CS kernel. Among multiple extractions of CS kernel we select, as for comparison, the extractions ART23 \cite{Moos:2023yfa}  (based only on DY data), ART25~\cite{Moos:2025sal} (based on DY and SIDIS data), and by ref.~\cite{Kang:2024dja} (based on the analogous analyses of EEC). The comparison is shown in fig.~\ref{fig:CS} (left). In general, the CS kernel obtained in the present fit has smaller values in comparison to all previous extractions, with very small uncertainty band. 

The disagreement between different extractions of CS kernel may indicate a tension between theory or experimental data, but most probably it indicates a problem with the method of analysis. It is also evident from the fits with fixed values of CS kernel, which produce a similarly good $\chi^2$. Therefore, we conclude that the present method of estimation of uncertainties is inadequate for these data set, and it \textit{extremely underestimates the uncertainty band}.

\begin{figure}[t]
\centering
\includegraphics[width=0.495\textwidth,valign=t]{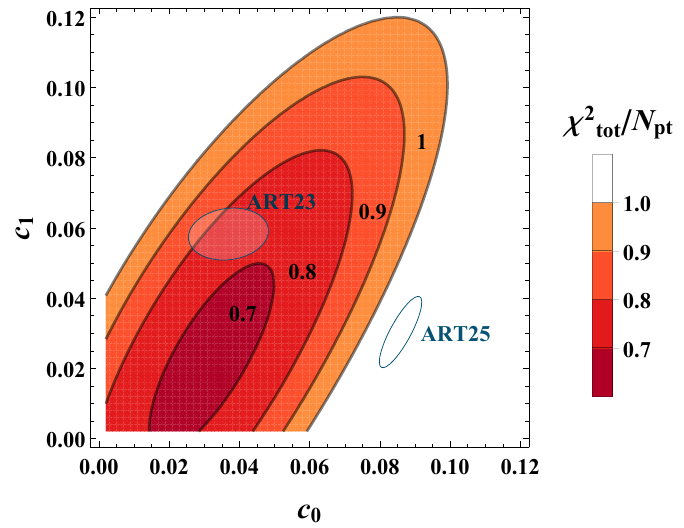}
\includegraphics[width=0.4\textwidth,valign=t]{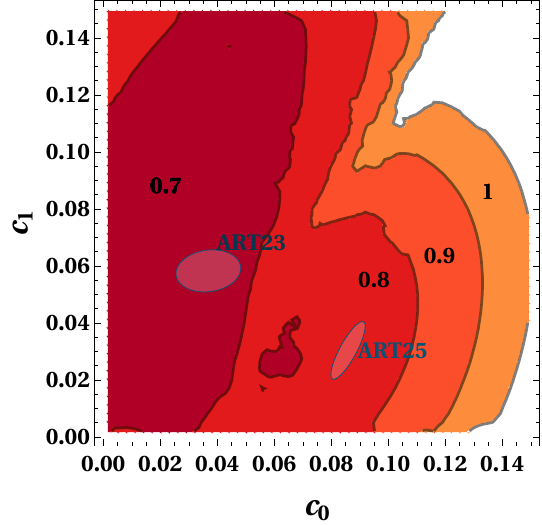}
\caption{\label{fig:c0c1-scan} Profile of the $\chi^2/N_{pt}$ in the plane of $c_{0,1}$ (with other model parameter being minimized for each point). The left(right) plot corresponds to the model 1(2).}
\end{figure}

Most probably this result is due to the fact that the errorbars for EEC are dominated by systematic sources. In this case many data sets have points distributed with a variation smaller than the announced uncertainty (i.e. they are very correlated). In other words, data points are ``aligned'' to some smooth curve. As a result, the distribution of points does not follow a Gaussian statistics with the corresponding deviation. 

In turn it leads to an underestimation of uncertainties in the Monte-Carlo replica method. This method propagates the uncertainty though the generation of pseudo-data based on the original data distribution \cite{Ball:2008by}. The produced pseudo-data have points distributed in the range of $2\sigma$, because the original data has $1\sigma$ distribution, which is amplified by the re-distribution to $2\sigma$. Correspondingly $\chi^2_{rep}$ generally converge to $\chi_{rep}^2/N\sim 2$. The minima of $\chi^2_{rep}$ are distributed with $\delta\chi^2\sim 1$, as it is required by the error-propagation. In the present case, the original data is practically noisy-less, and central values are distributed with some smaller variation (in the extreme case with $0\sigma$). Consequently, the pseudo-data is distributed with variation smaller that $2\sigma$ (in the extreme case $1\sigma$). The minimum of $\chi_{rep}^2$ for such data is almost the same as the minimum for the original data, which leads to an underestimation of the uncertainties.

In an attempt to estimate uncertainties more reliably, we used the assumption that $\chi^2/N_{pt}\sim 1$ indicates a good agreement with the data. This assumption is doubtful given obvious problems with statistical interpretation of the data. Nonetheless, it produces a visually adequate picture. 

We have performed the following test of the uncertainty of parameters. Fixing the values of $c_0$ and $c_1$ we fit the rest NP parameters and find the best $\chi^2/N_{pt}$. Then we plot the isolines of $\chi^2$ in the plane of $c_{0,1}$, which tell us the region of parameters that describe the data with a given level of statistical precision. The results of this scans for model 1 and 2 are shown in fig.~\ref{fig:c0c1-scan}. The allowed region of parameters is enormous. The $\chi^2/N_{pt}<1$ can be reached for any minimally reasonable model for CS kernel. If we remove the data with $\chi^2/N_{pt}<0.5$ from the fit, the region still remains large (with boundary approximately at $\chi^2/N_{pt}\sim 0.8$ in fig.~\ref{fig:c0c1-scan}).

Let us note that the scan of the model 2 parameters clearly demonstrates the presence of two minima, see fig.~\ref{fig:c0c1-scan} (right). One closer to the ART23, and another closer to ART25 determinations of CS kernel. In general, the fit within model 2 is less stable, because the optimization algorithm becomes unstable in the region in-between minima. In following discussion we use only model 1. 

\begin{figure}[t]
\centering
\includegraphics[width=0.94\textwidth,valign=t]{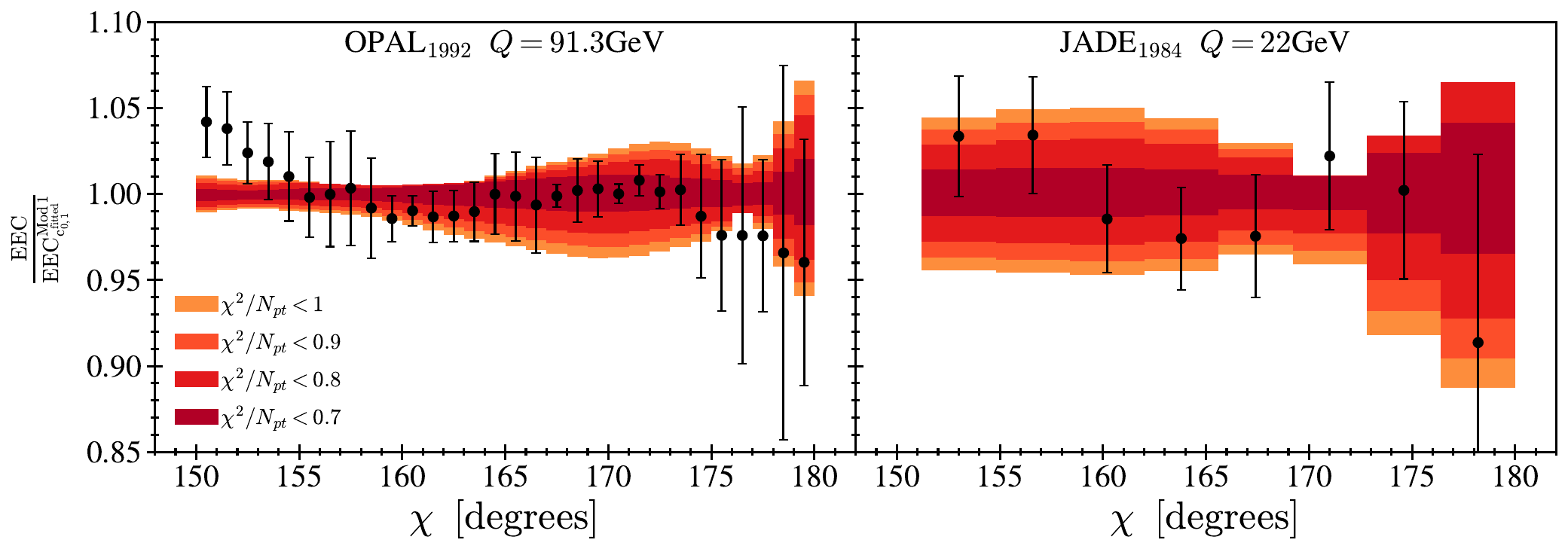}
\caption{\label{fig:c0c1-cross} Example of uncertainty band obtained with the fixed $\chi^2/N_{pt}$ value for OPAL and JADE experiments.}
\end{figure}

Note, that both $c_0$ and $c_1$ can reach the zero value but could not be smaller, since it is restricted by the positivity of CS kernel at large values of $b$. The case $c_0=c_1=0$ is not unphysical, because in this case our model of CS kernel eq.~(\ref{def:CS-kernel}) turn to a flat model with asymptotic value $\mathcal{D}_{\text{pert}}(B_{\text{NP}},\mu^*(B_{\text{NP}}))$. In this case the parameter $B_{\text{NP}}$ became the only NP parameter, and, in principle, should be varied. However, it is clear that such variation could not change the general conclusion.

In fig.~\ref{fig:CS}(right) we present the uncertainty band for the CS kernel corresponding to given values of $\chi^2/N_{pt}$. Within $\chi^2/N_{pt}<1$ the EEC data agrees with all extractions (we present ART23 and ART25, as examples, but most part of other determinations lie between these cases, see ref.~\cite{Moos:2025sal}). In fig.~\ref{fig:c0c1-cross} we demonstrate the inflation of the uncertainty band for the theoretical prediction in comparison with experimental data. The variation of the theory prediction with the variation of parameters is acceptable, in the sense that it is indeed compatible with the size of experimental uncertainty. It gives us confidence that this method provides an adequate estimation of uncertainties. Let us note, that even with a huge variation of non-perturbative parameters, the uncertainty band for a prediction is of the order of 1-5\%, and this implies that in order to get sensitivity to non-perturbative QCD physics, one should aim to 1\% experimental accuracy in vicinity of $\chi\sim 180^o$.

\begin{figure}[t]
\centering
\includegraphics[width=0.45\linewidth]{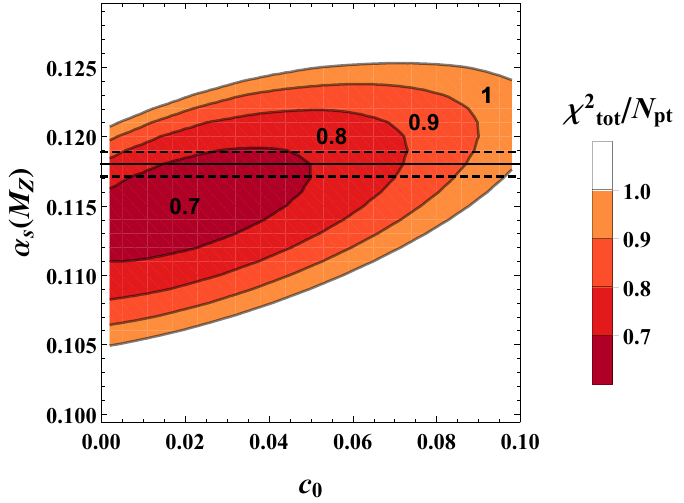}  
\caption{Profile of $\chi^2/N_{pt}$ in the plane of $\alpha_s$ and $c_0$ (with other nonperturbative parameters minimized for each point). The horizontal lines show the central value and uncertainty for the present determination of $\alpha_s(M_Z)$.}
\label{fig:scan-as}
\end{figure}

Recently, in ref.~\cite{Kang:2024dja}, it was argued that it is possible to determine the value of $\alpha_s$ using analyses of EEC in the TMD region. The authors of ref.~\cite{Kang:2024dja} implemented a similar model as here, and obtained $\alpha_s$ with competitively small uncertainties, using the uncertainty propagation with replica method. However, given our observation that this method does not produces adequate uncertainty band, this result is questionable. To cross-check this statement we have performed the scan of $\chi^2/N_{pt}$ versus $\alpha_s(M_Z)$ and $c_0$. The resulting iso-lines are shown in fig.~\ref{fig:scan-as}. 

Using the estimation of $\chi^2/N_{pt}$ as boundary for uncertainty we get the following results for $\alpha_s(M_Z)$
\begin{eqnarray}\nn
\chi^2/N_{pt}<1.0~: &\qquad & \alpha_s(M_Z)\in [0.105,0.125],
\\\label{alphas}
\chi^2/N_{pt}<0.9~: &\qquad & \alpha_s(M_Z)\in [0.106,0.124],
\\\nn
\chi^2/N_{pt}<0.8~: &\qquad & \alpha_s(M_Z)\in [0.108,0.120],
\\\nn
\chi^2/N_{pt}<0.7~: &\qquad & \alpha_s(M_Z)\in [0.111,0.119].
\end{eqnarray}
These estimations are not even compatible with the current uncertainty  for $\alpha_s(M_Z)=0.1180\pm0.0009$ \cite{ParticleDataGroup:2024cfk}. Therefore, we conclude that this kind of analyses and data are not capable of providing restrictions for $\alpha_s$.

\begin{figure}[t]
\centering
\includegraphics[width=0.96\textwidth]{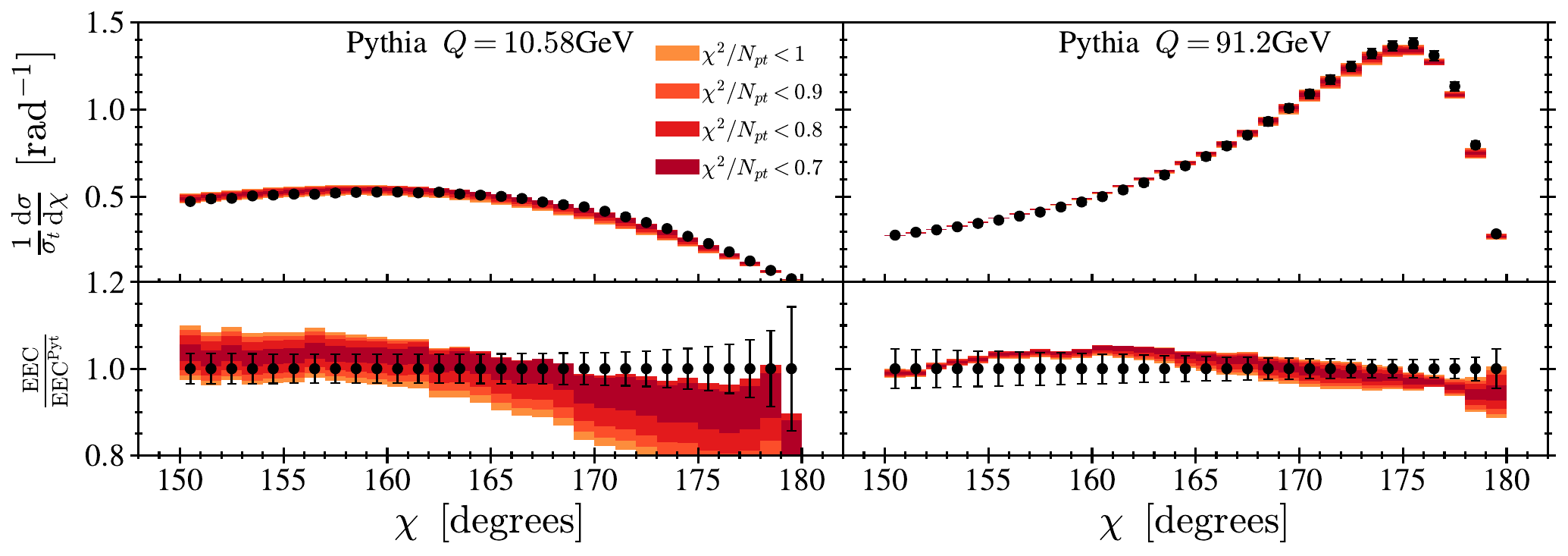}
\caption{Predictions for the back-to-back part of EEC for Belle and future facilities (FCCee) experiments based on this work (model 1). For comparison we demonstrate prediction of Pythia 8.314~\cite{Bierlich:2022pfr,RevModPhys.56.579}. The uncertainty band correspond to different levels of statistical agreement as discussed in the text.}
\label{fig:Belle}
\end{figure}

\section{Conclusions}
\label{sec:conclusions}

We have presented an analysis of the EEC in the back-to-back limit of $e^+e^-$ annihilation. In this regime, the EEC is described by the TMD factorization theorem with a jet function integrated over the momentum fraction. As a result, the nonperturbative physics is encapsulated in the Collins–Soper kernel and a single profile function. The perturbative component is taken at N$^4$LL accuracy, which includes N$^4$LL evolution and N$^3$LO matching.

The data used in this analysis come from experiments conducted in the 80's and 90's. We have tested the boundaries of the TMD factorization theorem, which is valid in the limit $\chi \to 180^\circ$, and found that this limit can be extended down to $\chi \sim 145^\circ$–$150^\circ$. This observation is in agreement with previous studies on the applicability of TMD factorization~\cite{Scimemi:2019cmh, Bacchetta:2022awv}. In total, we considered data from more than ten different experiments and selected 243 data points suitable for TMD analysis. Since many of these experiments do not provide a detailed breakdown of uncertainties, or report them indirectly, we have documented our own interpretation and treatment of the data in sec.~\ref{sec:data}. This represents one of the contributions of our work.

We found a remarkable agreement between theoretical predictions and the experimental data. However, this agreement must be interpreted with caution. In the absence of significant nonperturbative effects, the EEC is dominated by the tree-level expression and perturbative evolution, which aligns well with the observations. At the same time, the experimental uncertainties, which appear visually small, are effectively large due to limited experimental precision and unknown correlations. As a result, the sensitivity of the data to contributions from higher-order perturbative corrections and nonperturbative terms are negligible. Consequently, the data can be described with $\chi^2/N_{\text{pt}} < 1$ for any reasonable model of the Collins–Soper kernel, with a minimal tuning of the profile function.

Despite this, the uncertainty bands estimated from the data appear unrealistically narrow. This is an artifact of the Monte Carlo error-propagation procedure when applied to highly correlated data without specified correlation matrices. This observation strongly suggests that a significant portion of the data must be either discarded or reinterpreted in order to obtain a reliable estimate. Addressing this issue goes beyond the present study and is left for future work.

As an intermediate solution, we adopted the criterion of $\chi^2/N_{\text{pt}} < 1$ to estimate uncertainty bands. By performing a direct scan of the parameter space, we identified approximate uncertainty bands for the Collins–Soper kernel (see fig.~\ref{fig:CS}, right). These bands are broad enough to encompass nearly all existing models and extractions of the Collins–Soper kernel, despite significant disagreement between them. We also attempted to extract a constraint on the strong coupling constant $\alpha_s(M_Z)$, but the resulting bounds (summarized in eq.~(\ref{alphas})) are discouraging inconclusive.

We note that this study was inspired by ref.~\cite{Kang:2024dja}, and agrees with that work on theoretical aspects of factorization and many practical conclusions. However, the main disagreement concerns the estimation of uncertainties. Contrary to the conclusions of ref.~\cite{Kang:2024dja}, we assert that the current EEC data in the TMD-sensitive region are not sufficient to impose meaningful constraints on either the Collins–Soper kernel or the strong coupling constant. 

This situation could be improved by data from ongoing or future experiments, such as BELLE or future facilities (FCC-ee). In fig.~\ref{fig:Belle} we provide predictions for EEC at $Q=10.58$ GeV (BELLE case), and $Q=M_Z$ GeV (more suitable for the FCC-ee case). Obviously, the low-energy case is more suitable for the studies of nonperturbative QCD dynamics. Already 5\% precision measurement by BELLE will significantly restrict Collins-Soper kernel.
  
\acknowledgments

I.S. thanks G. Vita for discussion on this subject during his stage at CERN in summer 2024. A.V. is funded by the \textit{Atracci\'on de Talento Investigador} program of the Comunidad de Madrid (Spain) No. 2020-T1/TIC-20204.  
This project is supported by the Spanish Ministerio de Ciencias y Innovaci\'on Grant No. PID2022-136510NB-C31 funded by MCIN/AEI/ 10.13039/501100011033. This project has received funding from the European Union Horizon research Marie Skłodowska-Curie Actions – Staff Exchanges, HORIZON-MSCA-2023-SE-01-101182937-HeI.

\bibliography{bibFILEEC}
\end{document}